\documentclass[twocolumn,secnumarabic,amssymb, nobibnotes, aps, prc, superscriptaddress, nobalancelastpage,longbibliography,nofootinbib]{revtex4-1}

\usepackage{graphicx}
\usepackage{amsfonts}
\usepackage{amsmath,amssymb}
\usepackage{bm}
\usepackage{xcolor}
\usepackage{float}
\usepackage{dcolumn}
\usepackage{cancel}
\usepackage[symbol]{footmisc}
\usepackage{tcolorbox}
\usepackage{multirow}
\definecolor{lg}{gray}{0.95}

\renewcommand{\vec}[1]{\mbox{\boldmath $#1$}}

\begin{document}

\title{Isovector Monopole Resonance} 

\author{Xavier Roca-Maza}
\email{xavier.roca.maza@fqa.ub.es}
\affiliation{Departament de F\'isica Qu\`antica i Astrof\'isica, Mart\'i i Franqu\'es, 1, 08028 Barcelona, Spain}
\affiliation{Institut de Ci\`encies del Cosmos, Universitat de Barcelona, Mart\'i i Franqu\'es, 1, 08028 Barcelona, Spain}
\affiliation{Dipartimento di Fisica ``Aldo Pontremoli'', Universit\`a degli Studi di Milano, 20133 Milano, Italy}
\affiliation{INFN, Sezione di Milano, 20133 Milano, Italy}

\author{Gianluca Col\`o} 
\email{gianluca.colo@mi.infn.it}
\affiliation{Dipartimento di Fisica ``Aldo Pontremoli'', Universit\`a degli Studi di Milano, 20133 Milano, Italy}
\affiliation{INFN, Sezione di Milano, 20133 Milano, Italy}

\date{\today}
\begin{abstract}
We present a theoretical study of the three isospin channels of the isovector giant monopole resonance (IVGMR) within self-consistent Skyrme Hartree-Fock plus Random-Phase Approximation. A modified charge-exchange monopole operator is introduced to remove the dominant isobaric-analog-state contribution, thereby isolating the genuine radial isovector response. We derive the associated non-energy- and energy-weighted sum rules and establish approximate relations between the charge-exchange and non-charge-exchange strengths. The relative locations of the three IVGMR centroids are interpreted in terms of Coulomb displacement energies, the neutron-proton mass difference, and the finite-nucleus symmetry potential. Calculations for \(^{48}\)Ca, \(^{90}\)Zr, and \(^{208}\)Pb show that the non-charge-exchange centroid is accurately approximated by the average of the two charge-exchange centroids in medium-heavy and heavy nuclei. Using a broad set of Skyrme energy-density functionals, we find that the centroid energies of the three channels display a weak model dependence for a fixed nucleus, whereas the non-charge-exchange monopole polarizability exhibits a clear sensitivity to the density dependence of the symmetry energy. These results clarify the isospin structure of the IVGMR and support its use as a complementary probe of isovector nuclear properties.
\end{abstract}
\maketitle

\section{Introduction}
\label{intro}

Giant resonances in atomic nuclei are fundamental collective excitation modes that provide key insights into nuclear structure and dynamics \cite{Roca-Maza2018,hnp}. While the isoscalar giant monopole resonance (ISGMR) has been extensively studied and used as a probe of nuclear incompressibility and the equation of state \cite{Gar2018}, the isovector giant monopole resonance (IVGMR) - characterized by out-of-phase oscillations of neutron and proton densities - has received, so far, far less experimental attention \cite{Zeg2020}. The excitation energy of the non-charge exchange channel $\Delta T_z=0$ of the IVGMR has been accessed indirectly by averaging the excitation energies of the $\Delta T_z=\pm 1$ chanels obtained in charge-exchange reaction experiments (see, e.g., Ref.~\cite{bowman1987}). 

Theoretically, the non-charge-exchange IVGMR is expected to offer valuable and complementary information to its isoscalar counterpart. As all isovector resonances, its excitation energy is expected to be sensitive to both the volume and surface terms of the nuclear symmetry energy, making it relevant for constraining the properties of asymmetric nuclear matter and neutron-rich systems \cite{Oertel2017, Roca-Maza2018, Sorensen2024}. Moreover, the total non-energy-weighted sum rule of the charge-exchange IVGMR is linked to the difference between proton and neutron radii in nuclei, and thus provides an additional observable to probe the density dependence of the symmetry energy (cf. \cite{Zeg2020} and references therein).

Experimentally, identifying the IVGMR has proven challenging. Early evidence for the IVGMR came from pion charge-exchange reactions \cite{Erell84,Ere86,Irom86}, though the strength distribution was broad (10–15 MeV wide) and often obscured by non-resonant background contributions. Other techniques using heavy-ion reactions, such as the ($^7\text{Li}, ^7\text{Be} + \gamma$) reaction \cite{Nakayama1999}, allowed some access to non–spin-flip charge-exchange channels ($\Delta S=0$), but difficulties in background subtraction and angular momentum decomposition limited the possibility to interpret the data. 

A significant breakthrough came from the experiment by Scott et al. \cite{Sco2017}, who used the ($^{10}\text{Be}$, $^{10}\text{B}^*[1.74\,\text{MeV}]$) charge-exchange reaction at 100 AMeV on ${}^{28}$Si. This novel approach employed a secondary beam of $^{10}$Be and a $\gamma$-ray detection technique to selectively isolate $\Delta S=0$ transitions. The subsequent multipole decomposition analysis (MDA) of the angular distributions enabled the clear separation of IVGMR and IVGDR components. 

These developments establish the use of secondary radioactive ion beams and $\gamma$-tagged charge-exchange reactions as a powerful new method for studying the IVGMR. Furthermore, plans to increase the $^{10}$Be beam intensity may open the door to systematic studies of the IVGMR in heavier nuclei \cite{Zegers2023}.

In this context, the present work focuses on the theoretical study of the three isospin channels of the IVGMR, motivated by the recent experimental advancements. We aim to further explore the characteristics of the IVGMR—including its excitation energy, sum rules and sensitivity to isospin asymmetry—and assess its usefulness as a probe of nuclear structure and the density dependence of the symmetry energy.

The manuscript is organized as follows. In Sec.~\ref{theo} we will introduce the basic formalism and discuss the excitation operator, excitation energy and sum rules of the IVGMR. In Sec.~\ref{results} we will present our results and the conclusions will be laid in Sec.~\ref{conclusions}. 

\section{Theory}
\label{theo}
The IVGMR will be calculated through the Skyrme Hartree-Fock (HF) plus Random Phase Approximation (RPA) \cite{sch2019}. The Skyrme HF approach is one of the most successful techniques for the study of the ground-state properties of nuclei throughout the whole nuclear chart \cite{sch2019,Ben2003}. The small amplitude limit of time-dependent HF calculations, or RPA, allowed the description of many kinds of nuclear collective motion \cite{Paa2007}. In addition, the versatility of the Skyrme ansatz allows its use in more elaborate theoretical frameworks. The code employed for the calculations can be found in \cite{Col2021} for non-charge exchange resonances while the charge-exchange version is based on the same code (see, e.g. Ref.~\cite{Fra2005} for details). This theoretical framework is known to lie within the nuclear Energy Density Functional (EDF) approach.    

The main output from the calculations is the so called strength function,
\begin{equation}
S(E)=\sum_\nu \vert\langle \nu\vert\mathcal{O}\vert 0\rangle \vert^2\delta(E-E_\nu)
\end{equation}
where $\mathcal{O}$ is the excitation operator, $\vert 0\rangle$ is the nuclear ground-state and $\vert\nu\rangle$  labels an excited state of the nucleus. Another quantity that we will be discussing are the moments of the strength function, defined as  
\begin{equation}
m_\mu (k) = \int S(E) E^k dE\ .  
\label{moments}
\end{equation}
where $\mu=\pm$ refers to the charge-exchange $\Delta T_z = \pm 1$ channel, respectively, and $\mu=z$ refers to the non charge-exchange mode with $\Delta T_z = 0$.   

\begin{figure}[t!]
  \includegraphics[width=\linewidth]{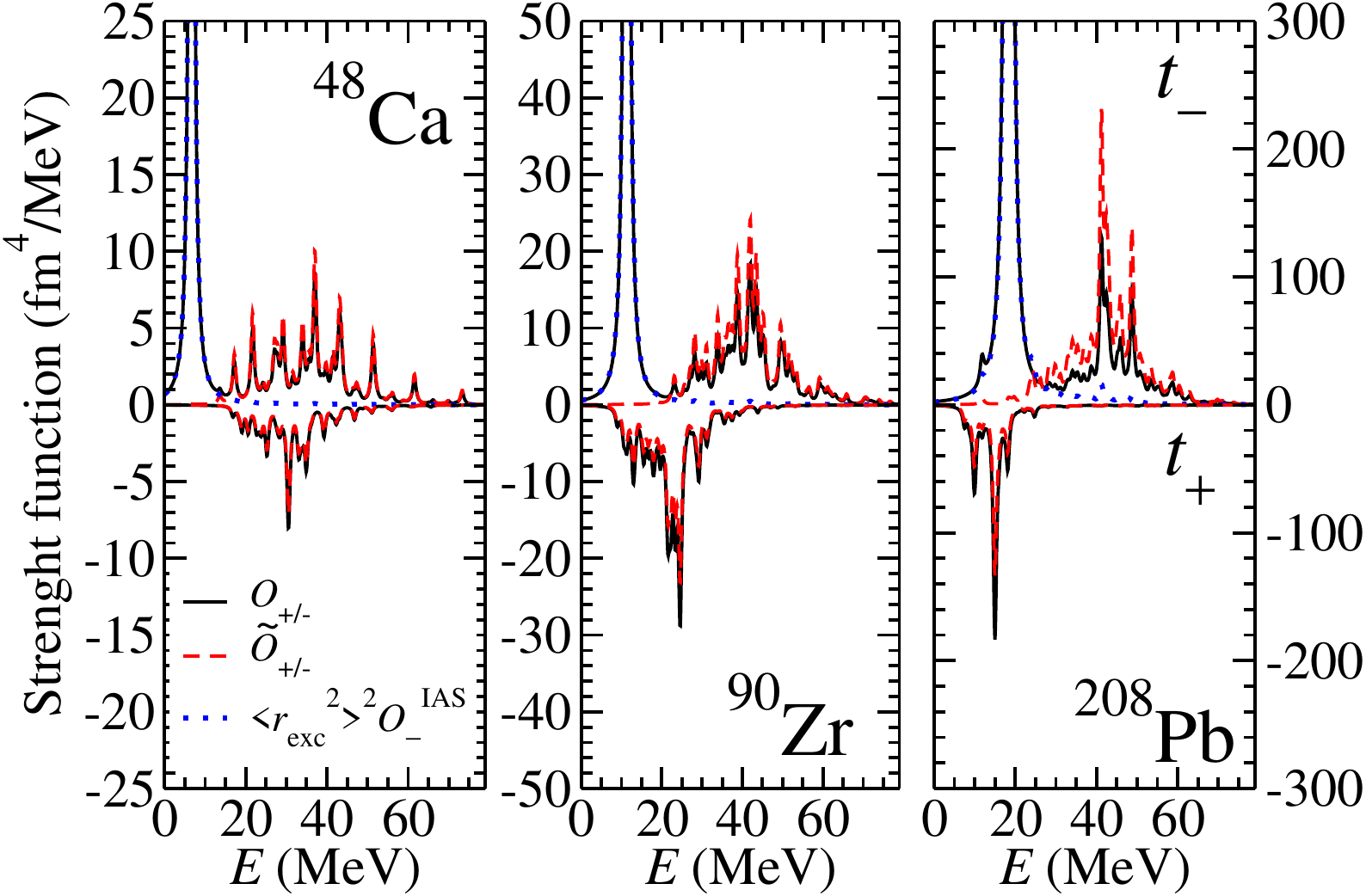}
  \caption{Strength funciton 
  (convoluted with Lorentzian functions having 1 MeV width) for the $t_-$ channels and $t_+$ channel (multiplied by $-1$) of the IVGMR and IAS in ${}^{48}$Ca (left panel), ${}^{90}$Zr (central panel) and ${}^{208}$Pb (right panel) as predicted by SAMi \cite{RocaMaza2012}. The employed operators $\mathcal{O}_\pm$, $\tilde{\mathcal{O}}_\pm$ and $\mathcal{O}^{\rm IAS}_-$, rescaled by $\langle r_{\rm exc}^2\rangle^2$, are detailed in the legend of the figure for the sake of clarity and also defined in the text. The $t_+$ channel results are multiplied by $-1$ such that all information can be given in a single figure. \label{fig1}}
\end{figure}

\subsection{Excitation operator}
\label{theo_ex}
The charge-exchange IVGMR operator has been defined in the literature as \cite{Auerbach83}: $\sum_{i=1}^A r_i^2\tau_\pm(i)$ where $\tau$ are the Pauli matrices in isospin space and relate to $t_\mu=\frac{1}{2}\tau_\mu$. An equivalent formulation used along the present work is $\mathcal{O}_\pm \equiv\sum_{i=1}^A r_i^2 Y_{00} t_\mu(i)$ which differs from the previous definition of the operator by a constant factor. 
The strength of these two operators, however, contain a large contribution from the Isobaric Analog State (IAS) which is excited by the operator $\mathcal{O}_\pm^{\rm IAS} = \sum_{i=1}^A t_\pm(i)$. 

In order to genuinely study the IVGMR one must redefine such operator to remove the effect of the IAS. For that one may proceed as follows. Knowing that the $t_-$ channel of the IAS exhaust almost the full non energy weighted sum rule (NEXSR): $m_-(0)-m_+(0)=N-Z$; in one single peak, the transition density of such peak can be written as proportional to the neutron excess. That is $\delta\rho_{\rm IAS} \propto \rho_{\rm exc}$ with $\rho_{\rm exc} \equiv \sum_i \vert\phi_i\vert^2$ with $i$ running on single-particle states belonging to the neutron excess. This allows to define a new IVGMR operator such that the IAS contribution to the IVGMR strength is zero,
\begin{equation}
\int d{\bm r} \delta\rho_{IAS} \tilde{\mathcal{O}}_\pm = 0
\end{equation}  
with $\tilde{\mathcal{O}}_\pm = \sum_{i=1}^A\left(r_i^2 - \xi\right)Y_{00} t_\pm(i)$. Such a condition implies that
\begin{equation}
\xi=\langle r_{\rm exc}^2\rangle \equiv \frac{1}{N-Z}\int d{\bm r} r^2\rho_{\rm exc}
\end{equation}
In Fig.~\ref{fig1}, the strength funciton (convoluted with Lorentzian functions having 1 MeV width) for the $t_{\pm}$ channels of the IVGMR operators $\mathcal{O}_\pm$ and $\tilde{\mathcal{O}}_\pm$ as well as the $t_-$ channel of the IAS multiplied by $\langle r_{\rm exc}^2\rangle^2$ in ${}^{48}$Ca, ${}^{90}$Zr and ${}^{208}$Pb as predicted by SAMi \cite{RocaMaza2012} are shown. All single EDF predictions will be given using the SAMi parametrization, unless otherwise stated. As it can be seen, the $\tilde{\mathcal{O}}_\pm$ (red dashed line) removes the effects of the IAS (blue dotted line) keeping the rest of the strength in the $t_-$ and $t_+$ channels almost unaltered. Thus, the new operators in the charge-exchange channel $\tilde{\mathcal{O}}_\pm$ will be adopted along this manuscript since they properly isolate the IVGMR resonance. 

For the case of the non charge-exchange IVGMR, a collective state excited by the operator $\mathcal{O}_z=\sum_{i=1}^A r_i^2Y_{00}t_z(i)$, the correction $\sum_{i=1}^A\langle r_{\rm exc}^2\rangle Y_{00}t_z(i)$ does not produce any effect on the nuclear response since $\langle r_{\rm exc}^2\rangle$ is a constant and $[\mathcal{H}, T_z]=0$ with $T_z\equiv\sum_{i=1}^A t_z(i)$. However, $\mathcal{O}_z$ may contain a center-of-mass contribution. In Appendix \ref{app2} we show that the correction goes as $1/A$ and that can be safely neglected for our purposes here and the studied nuclei.   
 
\begin{figure}[t!]
  \includegraphics[width=\linewidth]{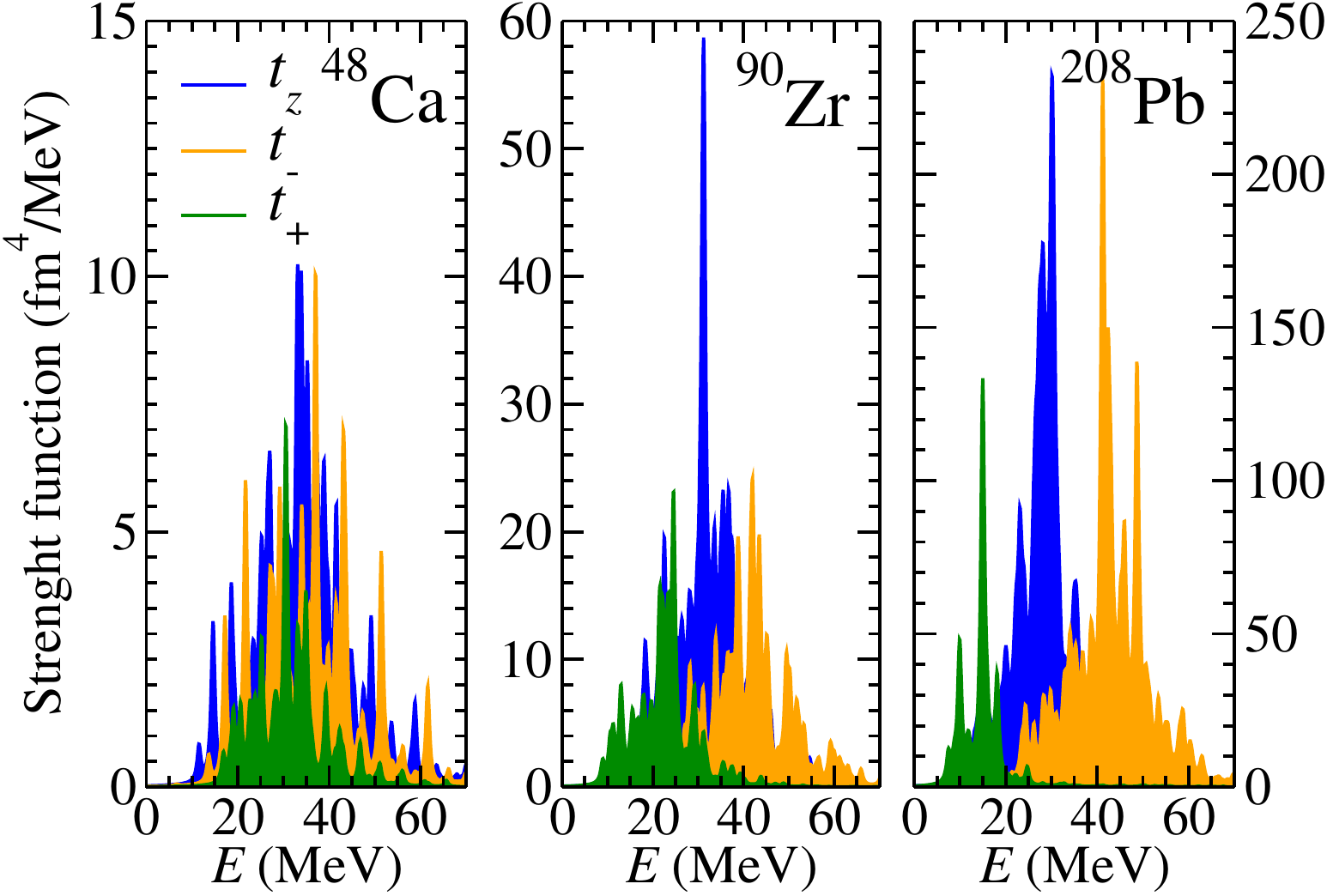}
  \caption{Strength funciton (convoluted with Lorentzian functions having 1 MeV width) for the $t_+$ channel (green), $t_z$ channel (blue) and $t_-$ channel (orange) of the IVGMR for ${}^{48}$Ca (left panel), ${}^{90}$Zr (central panel) and ${}^{208}$Pb (right panel) as predicted by SAMi \cite{RocaMaza2012} are shown. The employed operators are $\tilde{\mathcal{O}}_\pm$ and $\mathcal{O}_z$ as defined in the text.  \label{fig2}}
\end{figure}

For completness and to give an overall picture, the strength distributions as calculated with $\mathcal{O}_z$, together with those of the $t_\pm$ channels calculated with $\tilde{\mathcal{O}}_\pm$ are shown in Fig.~\ref{fig2}. The main peaks of the three chanels of the IVGMR in these three nuclei are more separated among them the heavier the nucleus is.

\subsection{Sum rules}
\label{theo_sr}

\begin{table}
\caption{NEWSR (fm$^4$) as calculated using the IVGMR $\mathcal{O}_z$, $\mathcal{O}_\pm$ and $\tilde{\mathcal{O}}_\pm$ operators for ${}^{48}$Ca, ${}^{90}$Zr and ${}^{208}$Pb. \label{tab1}}  
\begin{center}
\begin{tabular}{l|ccc} 
 \hline\hline
Operator& ${}^{48}$Ca & ${}^{90}$Zr&${}^{208}$Pb\\
\hline
$\tilde{\mathcal{O}}_-$ &100.926 &264.812 &1722.19  \\
$\mathcal{O}_-$         & 267.442& 639.926& 5970.7 \\
$\tilde{\mathcal{O}}_+$ &46.9135 &164.671 &427.169  \\
$\mathcal{O}_+$         & 53.3412& 207.236& 592.587 \\
$\mathcal{O}_z$         &145.731 &422.691 & 2019.84 \\
\hline\hline
\end{tabular}
\end{center}
\end{table}

Sum rules must be derived for the new operators $\tilde{\mathcal{O}}_\pm$ (cf. Ref.~\cite{Auerbach83} for the commonly used operators). 

\subsubsection{Non Energy Weighted Sum Rule}
\label{theo_sr_newsr}

We remind that $[t_+,t_-] = t_z$ and, thus,
\begin{equation}
  [\tilde{\mathcal{O}}_+, \tilde{\mathcal{O}}_-]=[\mathcal{O}_+, \mathcal{O}_-] + \frac{1}{4\pi}\sum_i^A t_z(i) \langle r_{\rm exc}^2\rangle^2\left(1-2\frac{r_i^2}{\langle r_{\rm exc}^2\rangle}\right), 
\end{equation}
where, as for the original IVGMR operators, $\tilde{\mathcal{O}}_-^\dag=\tilde{\mathcal{O}}_+$. The Non Energy Weighted Sum Rule (NEWSR) of the charge-exchange IVGMR can be then written as 
\begin{eqnarray}
  &&m_-(0)-m_+(0)=\langle 0 \vert[\tilde{\mathcal{O}}_+, \tilde{\mathcal{O}}_-]\vert 0\rangle\nonumber\\
  &&~~~~= \frac{1}{4\pi}\left(N\langle r_n^4\rangle-Z\langle r_p^4\rangle\right) \nonumber\\
  &&~~~~+\frac{N-Z}{4\pi}\langle r_{\rm exc}^2\rangle^2\left(1-2\frac{N\langle r_n^2\rangle-Z\langle r_p^2\rangle}{(N-Z)\langle r_{\rm exc}^2\rangle}\right) \ . 
\end{eqnarray}
This expression isolate the contribution of the IVGMR from that of the IAS to the NEWSR. The corresponding sum rule for the non corrected operators $\mathcal{O}_\pm$ is $\left(N\langle r_n^4\rangle-Z\langle r_p^4\rangle\right)/4\pi$ which is dominated by the IAS peak. 

For the studied cases, the NEWSR $m_-(0)$ is largely reduced by using the new operator: about 60-70\%; while the NEWSR $m_+(0)$ is only moderately reduced: about 10-30\% (cf. Table \ref{tab1}). For the case of the non-charge exchange channel, the NEWSR cannot be written analytically. However, it is interesting to see that $m_+(0)+m_-(0)\approx m_z(0)$\footnote{Note that within the standard operator definitions and assuming an isospin symmetric Hamiltonian (without Coulomb) $m_+(0)+m_-(0)=4m_z(0)$.}. Specifically, the accuracy of the relation is within 2\% in ${}^{48}$Ca and ${}^{90}$Zr and within a 7\% in ${}^{208}$Pb. This approximate relation can be written as a ground state expectation value of the involved operators as,
\begin{equation}
 \langle 0 \vert\{\tilde{\mathcal{O}}_+,\tilde{\mathcal{O}}_-\}\vert 0\rangle\approx \frac{1}{2}\langle 0 \vert\{\mathcal{O}_z^{\dag},\mathcal{O}_z\}\vert 0\rangle
\label{newsr_app} 
\end{equation}
within the Hartree-Fock approximation. Although no exact algebraic relation exists between the three NEWSRs, Eq.~(\ref{newsr_app}) could be understood as follows: the subtraction of the IAS removes the only large charge-exchange contribution with no analogue in the non-charge-exchange channel. Corrections to the latter equation will be due to differences on the proton and neutron radial wave functions due to Coulomb effects or different numbers of neutrons and protons; the deviation is indeed largest for \(^{208}\mathrm{Pb}\).

\subsubsection{Energy Weighted Sum Rule}
\label{theo_sr_ewsr}

We are also interested in the EWSR or $m_\mu(1)$ that can be written as
\begin{equation}
m_\mu(1) = \langle 0 \vert\mathcal{O}_\mu^\dag[\mathcal{H},\mathcal{O}_\mu]\vert 0\rangle
\end{equation}
For the charge-exchange chanels, since $\tilde{\mathcal{O}}_\pm^\dag\neq\tilde{\mathcal{O}}_\pm$, the latter expression cannot be written as a double commutator in the usual form. In order to obtain a double commutator, one must combine the $t_+$ and $t_-$ channels as follows 
\begin{eqnarray}
  &&m_+(1)+m_-(1) = \langle 0 \vert[\tilde{\mathcal{O}}_+,[\mathcal{H},\tilde{\mathcal{O}}_-]] \vert 0\rangle\nonumber\\
  &&~~~~=\langle 0 \vert[\mathcal{O}_+,[\mathcal{H},\mathcal{O}_-]] \vert 0\rangle\nonumber\\
  &&~~~~+\frac{1}{4\pi}\langle 0\vert\sum_{ij}\left(\langle r_{\rm exc}^2\rangle^2-2r_i^2\langle r_{\rm exc}^2\rangle\right) [t_+(i),[\mathcal{H},t_-(j)]]\vert 0\rangle\nonumber\\
  \label{dc}
\end{eqnarray}
where the first term in the r.h.s. coincides with the EWSR as derived in \cite{Auerbach83} and the second term is new [see Appendix \ref{app1} for a detailed derivation of Eq.~(\ref{dc})].

Analytic expresions assuming a Skyrme interaction can be found in Ref.~\cite{Auerbach83} for the $\mathcal{O}_\pm$ operators [cf. Eqs.~(61)-(65)]. The EWSR for the new operator $\tilde{\mathcal{O}}_\pm$ can be written, in a very similar fashion, in terms of different {\it enhancement} factors as
\begin{eqnarray}
m_+(1)+m_-(1) = \frac{1}{\pi}\frac{\hbar^2}{2m}A\langle r^2\rangle(1+\kappa+\eta_N+\tilde{\eta}_C)
\end{eqnarray} 
with  
\begin{eqnarray}
\kappa &\equiv& \frac{4}{A\langle r^2\rangle\rho_0}\left(\frac{m}{m_{IV}^*}-1\right)\int d{\bm r} r^2 \rho_n\rho_p\\
\eta_N &\equiv& \frac{1}{2A\langle r^2\rangle \rho_0}\left(\frac{m}{m_{IV}^*}-1\right)\int d{\bm r} r^2 \left(\rho_n-\rho_p\right)^2\\
\tilde{\eta}_C &\equiv& \frac{m}{2\hbar^2A\langle r^2\rangle}\int d{\bm r} \left(r^2-\langle r_{\rm exc}^2\rangle\right)^2\left(\rho_n-\rho_p\right)U_C, \nonumber \\  
\label{eta_c}
\end{eqnarray}
where $U_C$ is the Coulomb potential
\begin{equation}
U_C({\bm r}) = \int d{\bm r^\prime}\frac{e^2}{\vert{\bm r}-{\bm r^\prime}\vert}\rho_c({\bm r^\prime})
\end{equation}
and where a combination of the Skyrme parameters $t_1$, $x_1$, $t_2$ and $x_2$ have been written in terms of the isovector effective mass $m_{IV}^*$ calculated in infinite nuclear matter at saturation density $\rho_0$,
\begin{equation}
\frac{m_{IV}^*}{m} = \left\{1+\frac{m}{4\hbar^2}\rho_0\left[t_1(2+x_1+t_2(2+x_2))\right]\right\}^{-1}  \ .   
\label{eta_c_old}
\end{equation}
Only the last equation ($\tilde{\eta}_C$) has changed with respect to the results using $\mathcal{O}_\pm$,
\begin{eqnarray}
\eta_C &\equiv& \frac{m}{2\hbar^2A\langle r^2\rangle}\int d{\bm r} r^4\left(\rho_n-\rho_p\right)U_C,   
\label{eta_c_old}
\end{eqnarray}
clearly reducing the effect of this term to the total EWSR.

\begin{table}
\caption{Values of the enhancement factors $\kappa$, $\eta_N$, $\eta_C$ (corresponding to $\mathcal{O}_\pm$) and $\tilde{\eta}_C$ (corresponding to $\tilde{\mathcal{O}}_\pm$) of the different calculations shown in Table \ref{tab3}. \label{tab2}}  
\begin{center}
\begin{tabular}{l|ccc} 
 \hline\hline
Enhance. fac.& ${}^{48}$Ca & ${}^{90}$Zr&${}^{208}$Pb\\
\hline
$\kappa$         &0.2438& 0.2798&0.30071 \\
$\eta_N$        &0.0035& 0.0019&0.0063 \\
$\eta_C$        &0.3090 &0.4152 &2.0825 \\
$\tilde{\eta}_C$&0.0866& 0.1125& 0.58321\\
\hline\hline
\end{tabular}
\end{center}
\end{table}

We show in Table \ref{tab2} the values of the different enhancement factors. It is interesting to note that: i) $\eta_N$ is always negligible since it is proportional to the differences between neutron and proton densities (effect below 1\%); ii) $\kappa$ remain almost independent of the mass. It scales with $NZ/A^2$ which is 0.24 for ${}^{48}$Ca, 0.25 for ${}^{90}$Zr and 0.24 for ${}^{208}$Pb; while iii) $\tilde{\eta}_C$ increases with mass, being smaller than $\kappa$ for ${}^{48}$Ca and larger than $\kappa$ for ${}^{208}$Pb. Note that in all cases $\tilde{\eta}_C$ is about 4 times smaller than $\eta_C$.

The EWSR for the non-charge exchange IVGMR can be written as \cite{Auerbach83}, 
\begin{eqnarray}
m_z(1) = \frac{1}{\pi}\frac{\hbar^2}{2m}A\langle r^2\rangle(1+\kappa) \ . 
\end{eqnarray} 

In Table \ref{tab3} we show the EWSR as calculated using the IVGMR operators $\mathcal{O}_\pm$ and $\tilde{\mathcal{O}}_\pm$ as well as $\mathcal{O}_z$ operator. It is to be noted that the $m_z(1)$ is larger than $m_\pm(1)$ for ${}^{48}$Ca and ${}^{90}$Zr but not for ${}^{208}$Pb. This is due to the large increase of $\tilde{\eta}_C$ with respect to $\kappa$ in this nucleus.

\begin{table}
\caption{EWSR (MeV fm$^4$) as calculated using the IVGMR $\mathcal{O}_\pm$, $\tilde{\mathcal{O}}_\pm$ as well as $\mathcal{O}_z$ operator. \label{tab3}}  
\begin{center}
\begin{tabular}{l|ccc} 
 \hline\hline
Operator& ${}^{48}$Ca & ${}^{90}$Zr&${}^{208}$Pb\\
\hline
$\tilde{\mathcal{O}}_-$ &3702.64 & 11076.4& 73057.6 \\
$\mathcal{O}_-$         &4437.97 & 13466.7 & 134563.0 \\
$\tilde{\mathcal{O}}_+$ &1492.89 & 3734.75 &6682.66  \\
$\mathcal{O}_+$         &1694.11 & 4682.25& 9137.09 \\
$\mathcal{O}_z$         &4829.77&13561.3&54646.6 \\
 \hline\hline
\end{tabular}
\end{center}
\end{table}

The latter calculations allow us to give an estimate for the excitation energy of the IVGMR in the different channels and using the different operators. This is shown in Table \ref{tab5} as the centroid energy

\begin{equation}
E_c \equiv \frac{m_\mu(1)}{m_\mu(0)} \ .  
\end{equation}

\begin{table}
\caption{Centroid energies (MeV) as calculated using the IVM $\mathcal{O}_\pm$, $\tilde{\mathcal{O}}_\pm$ and $\mathcal{O}_z$ operator. \label{tab5}}  
\begin{center}
\begin{tabular}{l|ccc} 
 \hline\hline
Operator& ${}^{48}$Ca & ${}^{90}$Zr&${}^{208}$Pb\\
\hline
$\tilde{\mathcal{O}}_-$ &36.7 &41.8& 42.4\\
$\mathcal{O}_-$         &16.6 &21.0& 22.5\\
$\mathcal{O}_z$         &33.1 &32.1& 27.1\\
$\tilde{\mathcal{O}}_+$ &31.8 &22.7& 15.6\\
$\mathcal{O}_+$         &31.8 &22.6& 15.4\\
\hline\hline
\end{tabular}
\end{center}
\end{table}

It is worth noting that the splitting in the different channels, employing the operators $\tilde{\mathcal{O}}_\pm$ and $\mathcal{O}_z$, is not very far from being symmetric for all the three test cases shown here:
\begin{eqnarray}
  {}^{48}{\rm Ca}&& E_c(\tilde{\mathcal{O}}_-)-E_c(\mathcal{O}_z^{\rm CM}) = 3.6 {\rm ~MeV} \nonumber\\
  && E_c(\mathcal{O}_z^{\rm CM})-E_c(\tilde{\mathcal{O}}_+)= 1.3 {\rm ~MeV} \nonumber\\
  {}^{90}{\rm Zr}&& E_c(\tilde{\mathcal{O}}_-)-E_c(\mathcal{O}_z^{\rm CM}) = 9.7 {\rm ~MeV}\nonumber\\
  && E_c(\mathcal{O}_z^{\rm CM})-E_c(\tilde{\mathcal{O}}_+)= 9.4 {\rm ~MeV}\nonumber\\
  {}^{208}{\rm Pb}&&E_c(\tilde{\mathcal{O}}_-)-E_c(\mathcal{O}_z^{\rm CM}) = 15.3 {\rm ~MeV}\nonumber\\
  &&E_c(\mathcal{O}_z^{\rm CM})-E_c(\tilde{\mathcal{O}}_+)= 11.5 {\rm ~MeV}\ ; \nonumber
\end{eqnarray}
that is (cf. also \cite{Ere86}), 
\begin{eqnarray}
E_c(\mathcal{O}_z^{\rm CM}) &\approx& \frac{1}{2}\left [E_c(\tilde{\mathcal{O}}_+)+E_c(\tilde{\mathcal{O}}_-)\right] 
\label{const_app}
\end{eqnarray}
with an accuracy of 3\% in ${}^{48}$Ca, 0.5\% in ${}^{90}$Zr and below 7\% in ${}^{208}$Pb. To better understand the obtained relative energies of the three IVGMR channels, we will adopt a simple qualitative model in what follows \cite{Ere86, Leonardi1976, Auerbach83}. 

\subsubsection{Isospin energy splittings within a simple model}
\label{toy_model}

The three IVGMR channels can be discussed at the level of their centroid energies by exploiting the fact that the excitation operator carries one unit of isospin. Let the ground state of the target nucleus have isospin $T=T_z=(N-Z)/2$, and let \(P_{T'}\) project onto final states with total isospin \(T'=T-1,T,T+1\). 
For an isovector monopole operator \(\hat O_\mu\), the centroid in a given isospin channel is defined as \cite{Leonardi1976}

\begin{equation}
E_{T'}=
\frac{
\langle 0,T|
\hat O_\mu^\dagger P_{T'}\mathcal{H} P_{T'}\hat O_\mu
|0,T\rangle
}{
\langle 0,T|
\hat O_\mu^\dagger P_{T'}\hat O_\mu
|0,T\rangle
}~.
\label{30}
\end{equation}

The physical charge-exchange and non-charge-exchange responses select different combinations of the \(T-1\), \(T\), and \(T+1\) components; their identification with the \(t_+\), \(t_z\), and \(t_-\) channels follows the isospin convention adopted for the operators. Following Leonardi \cite{Leonardi1976}, the splitting of the three centroids can be parameterized by an isovector energy \(U_v\) and an isotensor energy \(U_t\). Defining

\begin{equation}
\Delta E_+ \equiv E_{T+1}-E_T,
\qquad
\Delta E_- \equiv E_T-E_{T-1},
\label{31}
\end{equation}

one obtains the exact relations

\begin{equation}
U_v=
-\left[
\frac{(2T+3)\Delta E_+}{2(T+1)(2T+1)}
+
\frac{(2T-1)\Delta E_-}{2T(2T+1)}
\right]~,
\label{32}
\end{equation}

and

\begin{equation}
U_t=
-\left[
\frac{\Delta E_+}{2(T+1)}
-
\frac{\Delta E_-}{2T}
\right]~.
\label{33}
\end{equation}

The quantity \(U_v\) determines the leading isospin dependence of the collective excitation, whereas \(U_t\) measures the departure from a purely isovector splitting. In the limit \(|U_t|\ll |U_v|\), Eqs.~(\ref{32}) and (\ref{33}) give

\begin{equation}
\Delta E_+ \simeq -(T+1)U_v,
\qquad
\Delta E_- \simeq -T U_v \ .
\label{24}
\end{equation}
This estimate neglects the small difference between the neutron and proton masses $\delta m_{np}\equiv m_n-m_p\approx 1.3$ MeV as well as the contribution from the Coulomb interaction, the Coulomb displacement energy $\Delta E_C$, that can be approximately estimated as, 
\begin{eqnarray}
\Delta E_C^{(+)} &\simeq& E_C(Z-1)-E_C(Z)\nonumber\\
\Delta E_C^{(-)} &\simeq& E_C(Z) - E_C(Z+1)\nonumber
\end{eqnarray} 
where $E_C(Z)$ is the Coulomb energy associated to the nucleus with $Z$ protons. One may assume $\Delta E_C^{(+)}\simeq \Delta E_C^{(-)}\equiv\Delta E_C$ within our simple model here. To take into account these two other effects one should add them to Eq.~(\ref{24}) as  
\begin{eqnarray}
\Delta E_\pm&\simeq& \Delta E_C\mp\delta m_{np}-\frac{N-Z}{2}U_v 
\end{eqnarray} 
where we have considered a neutron-rich heavy nucleus, for which \(T\gg 1\). Therefore, the centroid of the intermediate \(T\) component is then expected to be close to the average of the two outer components.

This model justifies  our RPA results in a simple and qualitative manner.  It is thus a consequence of isospin geometry and of a small isotensor contribution; it does not imply that the symmetry energy is negligible. On the contrary, the symmetry potential is contained in the leading isovector coefficient \(U_v\), while \(U_t\) quantifies the residual nonlinearity of the splitting.

Within this same model and assumptions, the term $U_v$ can be set to $U_v = V_1/(2A)$ as in \cite{lipparini1989} (cf. Sec.~9) where $V_1$ is the symmetry potential and can be written as $V_1 = 4a_A$ \cite{bm}, where $a_A=24$ MeV corresponds to symmetry energy coefficient in the semi-empirical mass formula. Within EDFs, this term has been predicted with very little dispersion due to the need of reproducing nuclear masses in the calibration procedure \cite{Centelles2009}. Estimating for the heaviest nucleus where this simple model better applies, one obtains ($\Delta E_C \sim 19$ MeV), 
\begin{eqnarray}
\
\Delta E_+({}^{208}{\rm Pb})&\simeq& 
13\ {\rm MeV}  \nonumber \\
\Delta E_-({}^{208}{\rm Pb})&\simeq& 
15\ {\rm MeV}  \nonumber 
\end{eqnarray} 

The microscopic Skyrme results can consequently be interpreted in a simple way. The separation of the three channels is controlled primarily by the isovector coefficient \(U_v\), after the appropriate Coulomb-displacement and neutron-proton-mass corrections have been made. The small isotensor coefficient \(U_t\) explains why the central channel remains close to the mean of the two charge-exchange centroids. Since Coulomb displacement energies are constrained by charge radii and the finite-nucleus symmetry potential by masses, both the channel splittings and the three centroids are expected to show only a weak dependence with the EDF model. This will be confirmed by a more systematic study using microscopic results later (cf. Sec.~\ref{results}).



\subsubsection{Brief summary on the NEWSR and EWSR}
Summarizing, we have found two approximate and three exact relations - within the Skyrme EDF framework - involving six moments of the strength function     
\begin{eqnarray} 
  \frac{m_+(1)}{m_+(0)} + \frac{m_-(1)}{m_-(0)}&\approx&2\frac{m_z(1)}{m_z(0)} \nonumber\\
  m_+(0)+m_-(0)&\approx&m_z(0)\nonumber\\
  m_-(0)-m_+(0)&=&\frac{1}{4\pi}\left(N\langle r_n^4\rangle-Z\langle r_p^4\rangle\right) \nonumber\\
  &+&\frac{N-Z}{4\pi}\langle r_{\rm exc}^2\rangle^2\left(1-2\frac{N\langle r_n^2\rangle-Z\langle r_p^2\rangle}{(N-Z)\langle r_{\rm exc}^2\rangle}\right) \nonumber\\ 
m_+(1)+m_-(1) &=& \frac{1}{\pi}\frac{\hbar^2}{2m}A\langle r^2\rangle(1+\kappa+\eta_N+\tilde{\eta}_C)\nonumber\\
  m_z(1) &=& \frac{4}{\pi}\frac{NZ}{A}\frac{\hbar^2}{2m}\langle r^2\rangle (1+\kappa) \ .    
\end{eqnarray}

\subsubsection{Inverse Energy Weighted Sum Rule}
\label{theo_sr_iewsr}

\begin{table}
\caption{IEWSR as calculated using the IVM $\mathcal{O}_\pm$, $\tilde{\mathcal{O}}_\pm$ and $\mathcal{O}_z$ operators. \label{tab6}}  
\begin{center}
\begin{tabular}{l|ccc} 
 \hline\hline
Operator& ${}^{48}$Ca & ${}^{90}$Zr&${}^{208}$Pb\\
\hline
$\tilde{\mathcal{O}}_-$ &3.20& 6.91& 44.34 \\
$\mathcal{O}_-$         &33.12& 47.3& 312.65  \\
$\tilde{\mathcal{O}}_+$ &1.65& 8.58& 33.13 \\
$\mathcal{O}_+$         &1.87& 10.85 & 46.38 \\
$\mathcal{O}_z$         &5.23& 14.78 & 87.61 \\
\hline\hline
\end{tabular}
\end{center}
\end{table}

The Inverse Energy Weighted Sum Rule (IEWSR), $m_\mu(-1)$, proportional to the static polarizability $\alpha_M = 2 m_\mu(-1)$ can be calculated within the RPA as in Eq.~(\ref{moments}) or by a constrained ground state calculation following the dielectric theorem \cite{bohigas1979, capelli2009, Hinohara2015},
\begin{eqnarray}
\frac{1}{m_\mu(-1)} &=& 2\frac{\partial^2 \langle\mathcal{H}\rangle}{\partial \langle \mathcal{O}_\mu\rangle^2}\Bigg\vert_{\lambda = 0} \ 
\label{dtheo}
\end{eqnarray}
where $\lambda$ is the parameter controlling the perturbation of the ground state $\mathcal{H} = \mathcal{H}_0 +\lambda\mathcal{O}_\mu$. The perturbation $\mathcal{O}_\mu$ in the present case would correspond to the IVGMR operator. 

In Table \ref{tab6}, the RPA results for the $m_\mu(-1)$ are shown. In Table \ref{tab7} the RPA results for the so called constrained energy
\begin{eqnarray}
E_{\rm con}\equiv \sqrt{\frac{m_\mu(1)}{m_\mu(-1)}}
\end{eqnarray}
are also displayed. As expected the constrained energy is equal or lower than the centroid energy and the differences between energies associated to the different channels are
\begin{eqnarray}
  {}^{48}{\rm Ca}&& E_{\rm con}(\tilde{\mathcal{O}}_-)-E_{\rm con}(\mathcal{O}_z^{\rm CM}) = 3.6 {\rm ~MeV} \nonumber\\
  && E_{\rm con}(\mathcal{O}_z^{\rm CM})-E_{\rm con}(\tilde{\mathcal{O}}_+)=  0.3 {\rm ~MeV} \nonumber\\
  {}^{90}{\rm Zr}&& E_{\rm con}(\tilde{\mathcal{O}}_-)-E_{\rm con}(\mathcal{O}_z^{\rm CM}) =  9.7 {\rm ~MeV}\nonumber\\
  && E_{\rm con}(\mathcal{O}_z^{\rm CM})-E_{\rm con}(\tilde{\mathcal{O}}_+)=  9.4 {\rm ~MeV}\nonumber\\
  {}^{208}{\rm Pb}&&E_{\rm con}(\tilde{\mathcal{O}}_-)-E_{\rm con}(\mathcal{O}_z^{\rm CM}) =  15.6 {\rm ~MeV}\nonumber\\
  &&E_{\rm con}(\mathcal{O}_z^{\rm CM})-E_{\rm con}(\tilde{\mathcal{O}}_+)= 10.8 {\rm ~MeV} \nonumber
\end{eqnarray}
keeping a similar relation as the centroid energies discussed above.

One of the advantages of estimating the same quantities using the dielectric theorem, Eq.~(\ref{dtheo}), is that assuming a simple model for the Hamiltonian, one can obtain analitic expressions in terms of the nuclear model parameters that can help understanding the relevant physics determining $m_\mu(-1)$.    

\begin{table}
\caption{Constrained energies as calculated using the IVM $\mathcal{O}_\pm$, $\tilde{\mathcal{O}}_\pm$ and $\mathcal{O}_z$ operators. \label{tab7}}  
\begin{center}
\begin{tabular}{l|ccc} 
 \hline\hline
Operator& ${}^{48}$Ca & ${}^{90}$Zr&${}^{208}$Pb\\
\hline
$\tilde{\mathcal{O}}_-$ &34.0& 40.0 &40.6    \\
$\mathcal{O}_-$         &11.6& 16.9 &20.7  \\
$\mathcal{O}_z$         &30.4& 30.3 &25.0  \\
$\tilde{\mathcal{O}}_+$ &30.1& 20.9 &14.2  \\
$\mathcal{O}_+$         &30.1& 20.8 &14.0  \\
\hline\hline
\end{tabular}
\end{center}
\end{table}

For the non-charge exchange excitation operator $\mathcal{O}_z$, the expectation value in the ground state of this operator goes as, $\langle \mathcal{O}_z\rangle 
 \sim \langle r_n^2\rangle-\langle r_p^2\rangle$. Hence a variation of the excitation operator is due to a variation of the isovector density,
\begin{eqnarray}
\langle r_n^2\rangle-\langle r_p^2\rangle = \int d{\bm r} r^2 \left[\rho_n({\bm r})-\rho_p({\bm r})\right] \ .
\end{eqnarray}
This implies that the IEWSR -proportional to the polarizability ($\alpha_M$) - must be proportional to the second derivative of the expectation value of the energy density with respect to the isovector density \cite{kirvine1980,casas1985,bowman1987}. That is, the symmetry energy of the finite nucleus.

We remind here that the dielctric theorem applied to the ISGMR predicts a relation with the finite nucleus incompressibility. This feature is completely analogous to what happens with the IV and IS Giant Dipole Resonances: while the former depends on the symmetry energy, the latter depends on the nucleus incompressibility. This result can be extended to any IV and IS multipole based on such general theorem.   

\begin{figure}[t!]
  \includegraphics[width=\linewidth]{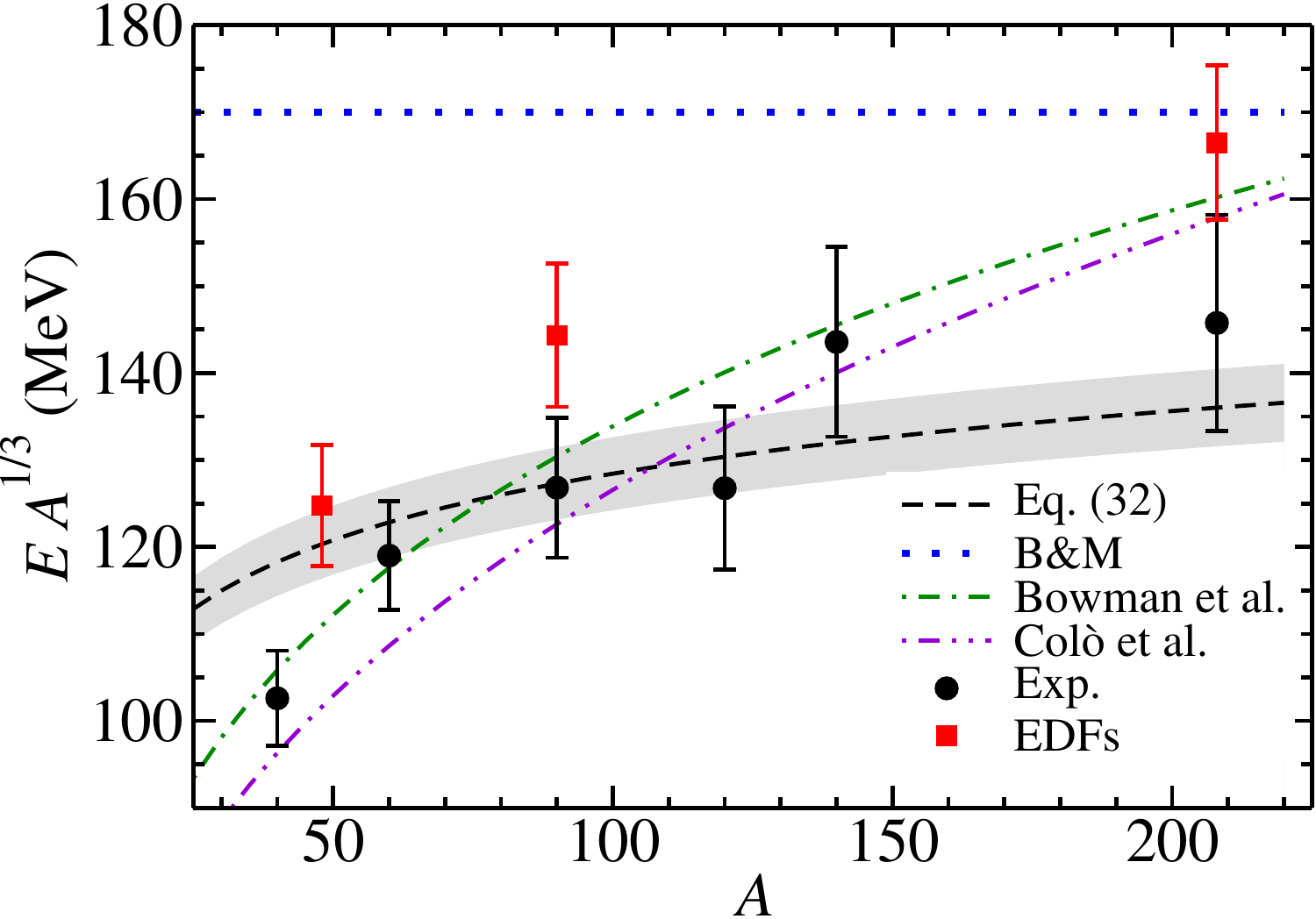}
  \caption{Excitation energies of the non-charge exchange IVGMR times $A^{1/3}$ as a function of the mass number $A$. Experimental data (Exp.) are taken from Ref.~\cite{Ere86} and plotted after applying the approximation in  Eq.~(\ref{const_app}). The red square symbols represent the average and estimated spread in the employed EDFs (see text for details). The dashed black line and shadowed grey band represent the pocket formula in Eq.~(\ref{pocket}). The green dash-dotted line corresponds to the model developed in \cite{bowman1987} that has a leading  $A^{-1/6}$ dependence instead of the $A^{-1/3}$ shown in Eq.~(\ref{pocket}). 
  The same dependence ($A^{-1/6}$) was derived in the model given in \cite{Colo1995} and shown here as dash-double-dotted violet line. \label{fig3bis}}
\end{figure}

In order to give a simple estimate of the monopole polarizability, one can use the Droplet Model \cite{myers1974}. In analogy with the dipole polarizability within the same approximation \cite{meyer1982} and similarly to the hydrodinamical model \cite{bowman1987}, one finds the following approximate expression,    
\begin{eqnarray}
\alpha_M &\approx& \frac{A}{4\pi}\frac{\langle r^4\rangle-\langle r^2\rangle^2}{J}\left[1+\frac{7}{3}\frac{L}{J}A^{-1/3}\right]
\end{eqnarray}
where $J$ and $L$ are the symmetry energy and its slope at nuclear matter saturation density.

Now we can give a simple estimate for the excitation energy of the non-charge-exchange IVGMR. That is, 
\begin{eqnarray}
  E_{\rm con}^2(\mathcal{O}_z^{\rm CM})&=&\frac{m_z(1)}{m_z(-1)}\nonumber\\
  &\approx&32\frac{NZ}{A^2}\frac{\hbar^2}{2m}\frac{\langle r^2\rangle}{\langle r^4\rangle-\langle r^2\rangle^2} \frac{(1+\kappa)J}{1+\frac{7}{3}\frac{L}{J}A^{-1/3}} \ . \nonumber\\
\end{eqnarray}
From the last equation one can derive a pocket formula for the excitation energy of the IVGMR by assuming a sharp sphere model for the nucleus density distribution,   
\begin{eqnarray}
E_{\rm con}^2(\mathcal{O}_z^{\rm CM})&\approx&70\frac{\hbar^2}{2m R^2} \frac{J}{1+\frac{7}{3}\frac{L}{J}A^{-1/3}} \nonumber\\
E_{\rm con}(\mathcal{O}_z^{\rm CM})&\approx& (182\pm 12) A^{-1/3}\left(1+\frac{14}{3}A^{-1/3}\right)^{-1/2}
\label{pocket}
\end{eqnarray}
where the result is in MeV, we have assumed $R=r_0 A^{1/3}$ with $r_0=1.14$ fm that corresponds to a saturation density of $\rho_0=0.16$ fm$^{-3}$, $\kappa \sim NZ/A^2\approx 1/4$, $J$ to be in the range $27.5-31.5$ MeV and $L/J$ have been set to 2. This formula corrects the well known Bohr and Mottelson formula $E = 170A^{-1/3}$ MeV \cite{bm} 
and it is different from the formulas given in \cite{bowman1987,Colo1995}; note that
in Ref.~\cite{Colo1995} the goal
was to understand the interaction dependence of the IVGMR energy and not its mass dependence.

In Fig.~\ref{fig3bis}, we show the results of these formulas together with experimental data and different microscopic calculations based on a large set of Skyrme EDFs. Specifically, we plot the excitation energies of the non-charge exchange IVGMR times $A^{1/3}$ as a function of the mass number $A$ determined experimentally (Exp.) in Ref.~\cite{Ere86}, with Eq.~(\ref{pocket}) and as predicted by the two models in Refs.~\cite{bowman1987, Colo1995}. As we detail in the following section, we use 40 Skyrme EDFs for the study of the IVGMR in ${}^{48}$Ca, ${}^{90}$Zr and ${}^{208}$Pb. The results shown in the present figure represent the average and spread of the model predictions (red squares). It is important to note that the experimental data have been obtained after applying the approximation in  Eq.~(\ref{const_app}) since the quantity determined in the experiment were the two charge exchange chanels for ${}^{40}$Ca, ${}^{60}$Ni, ${}^{90}$Zr, ${}^{120}$Sn, ${}^{140}$Ce and ${}^{208}$Pb. As we will discuss also below, the expression in Eq.~(\ref{const_app}) works better for medium-heavy to heavy nuclei with well separated peaks in the three isospin chanels.    

\section{Results}
\label{results}

\subsection{${}^{208}$Pb}

\begin{figure}[t!]
  \includegraphics[width=\linewidth]{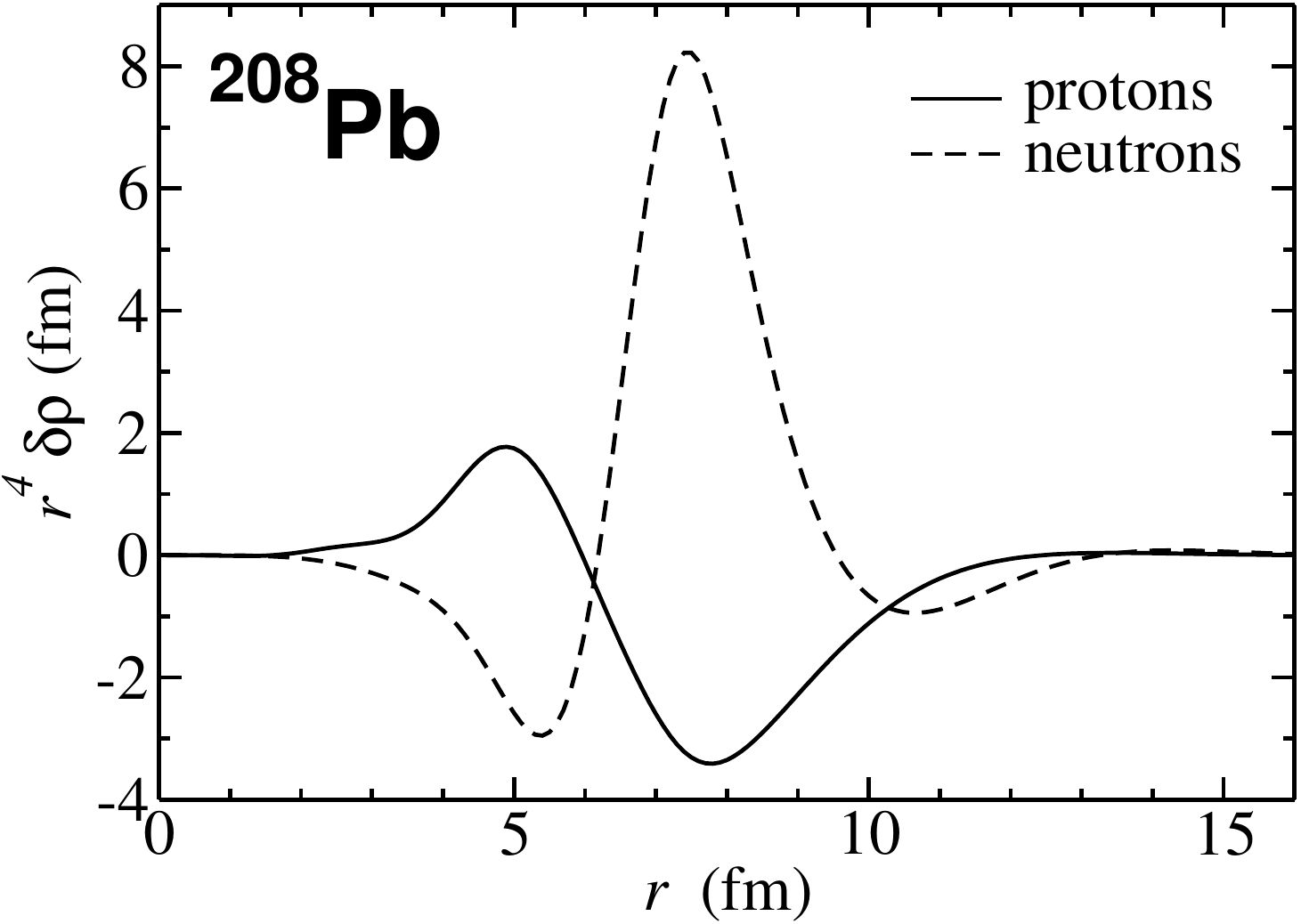}
  \caption{Neutron and proton transition densities for the most significant state at 30 MeV, expressed as a function of radial distance from the center, as predicted by SAMi for ${}^{208}$Pb. \label{fig4}}
\end{figure}

In this section we will present our results on the three isospin chanels of the IVGMR for ${}^{208}$Pb. First of all, in Fig.~\ref{fig4} we show the transition density multiplied by $r^4$ of the most important peak of the non-charge exchange IVGMR at about 30 MeV\footnote{Keeping in mind that the strength associated to such peak will correspond to the radial integral of the difference of the two curves given in this figure.} as predicted by the SAMi EDF (see also Fig.~\ref{fig2}). This is shown as an example of the clear isovector behavior of the resonance peak, as expected from a collective isovector mode.

\begin{figure}[t!]
  \includegraphics[width=\linewidth]{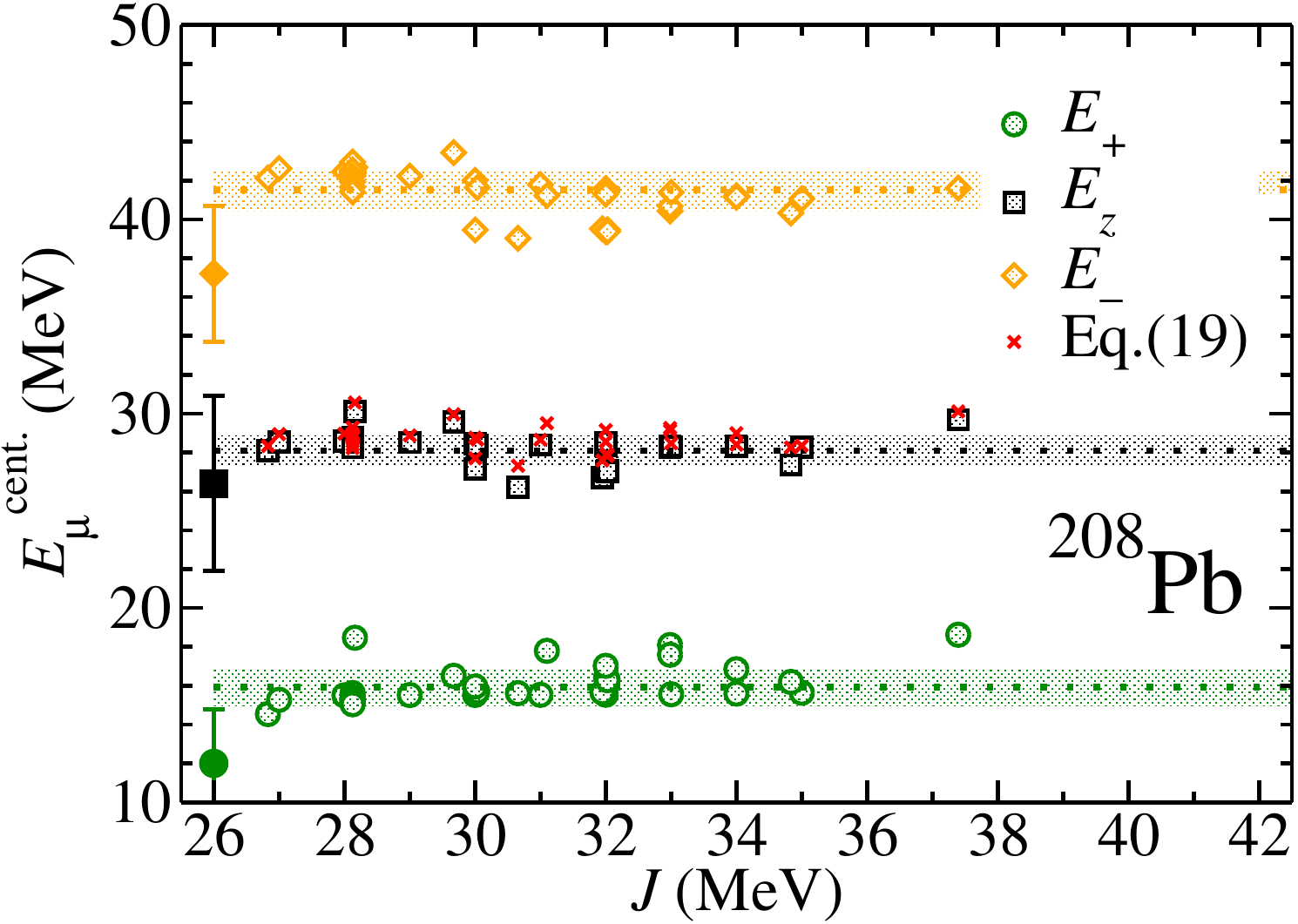}
  \caption{Centroid excitation energy of the IVGMR in the three isospin chanels  as a function of the symmetry energy at saturation density and as predicted by a set of 40 Skyrme EDFs. Data has been taken from \cite{Ere86}. \label{fig5}}
\end{figure}

In Fig.~\ref{fig5}, the centroid energies ($m_\mu(1)/m_\mu(0)$) in the three isospin chanels are depicted as a function of the symmetry energy. To test the model dependence we have used 40 different Skyrme EDFs with different properties of the symmetry energy, effective masses and nuclear matter incompressibility among other global properties. The models are: SAMi \cite{RocaMaza2012}, SAMi-T \cite{Shen2019} as well as the SAMi families of systematically varied interactions SAMi-J, SAMi-K and SAMi-m$^*$ \cite{Roca-Maza2013}; and some Skyrme EDFs of common use in the literature KDE0 and KDE0-J33 \cite{Agrawal2005}, SGII \cite{Giai1981}, SIII \cite{Beiner1975}, SKI3 \cite{Reinhard1995}, SKM* \cite{Bartel1982}, SKO' \cite{Reinhard1999}, SLy5-T \cite{Colo2007}, SLy4 and SLy5 \cite{Chabanat1998}, SK255 \cite{Agrawal2003}, SV-sym32, SV-sym34 and SV-min \cite{kupfel2009}, SkP \cite{Dobaczewski1984}, and SkX \cite{Brown1998}. As previously discussed  in Sec.~\ref{toy_model}, the excitation energy of the IVGMR is expected to weakly depend on the EDF employed, specially for neutron-rich heavy nuclei and well callibrated models to nuclear masses and charge radii. This result is confirmed by our microscopic calculations. This further validates the relation given in Eq.~(\ref{const_app}) shown as red crosses in the figure and previously used in the literature (see, e.g., Refs.~\cite{bowman1987} or \cite{Ere86}).

\begin{figure*}[t!]
  \includegraphics[width=0.45\linewidth]{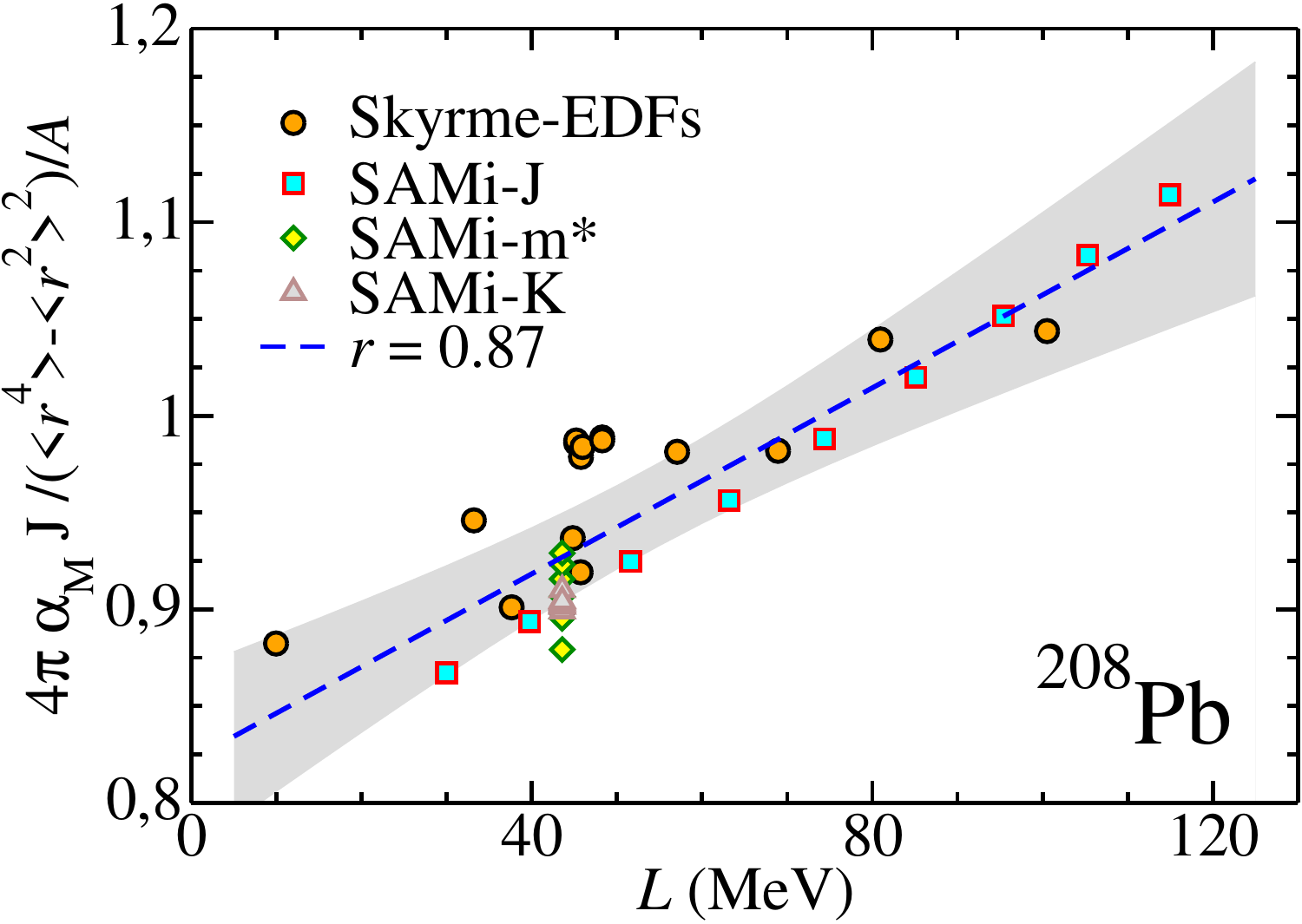}
  \hfill
  \includegraphics[width=0.45\linewidth]{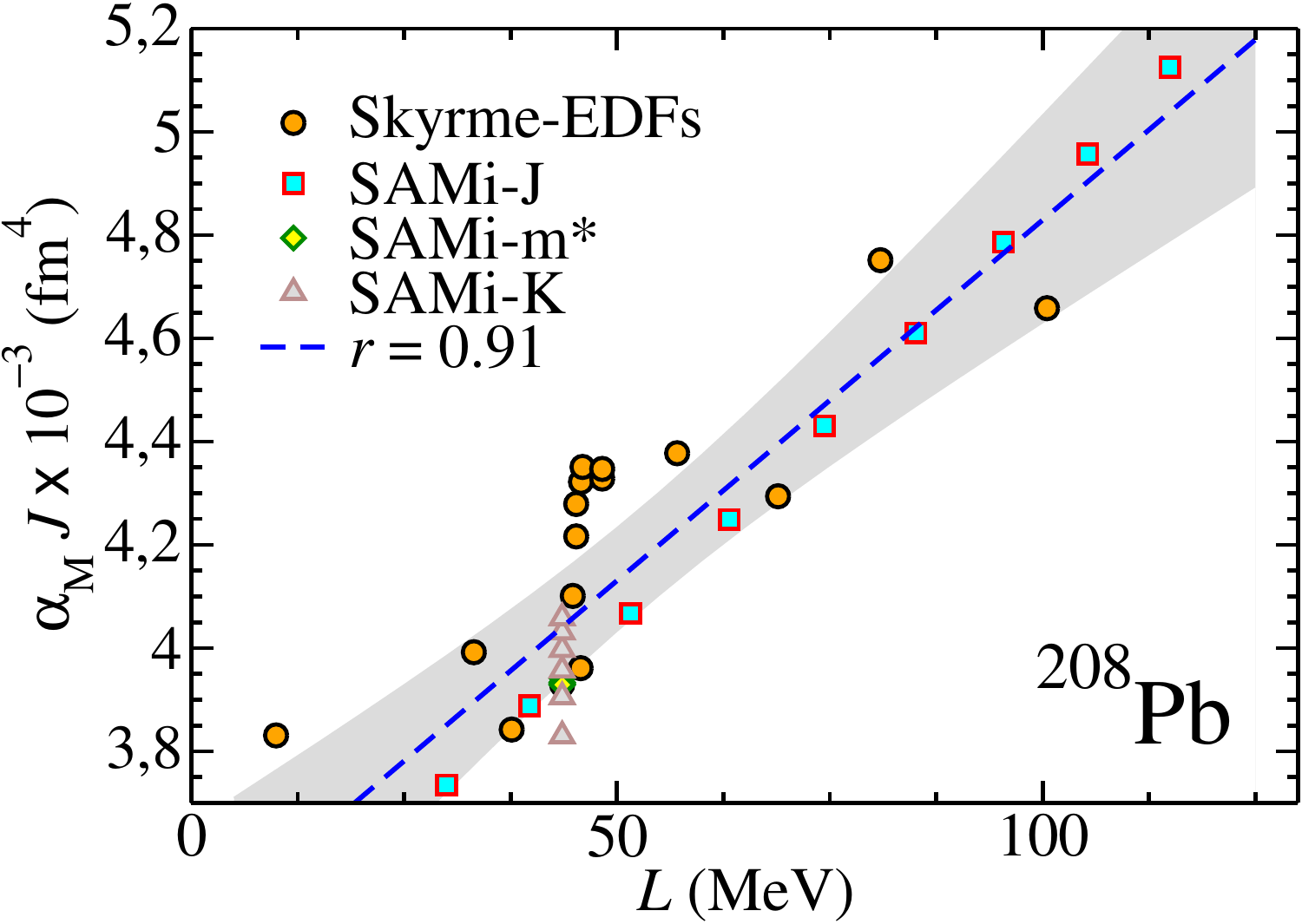}
  \caption{Non-charge-exchange isovector monopole  polarizability ($\alpha_M$) times the symmetry energy at saturation density as a function of the slope of the symetery energy at saturation density. The difference between the two panels is on the vertical axes (see text). Results have been predicted by a set of 40 Skyrme EDFs. See the text for details. The shaded region depicts the limits of the 95\% prediction band for the regression (see, e.g., Chap. 3 of Ref.~\cite{Skovgaard1998}).\label{fig6}}
\end{figure*}

By using the same models as in Fig.~\ref{fig5}, we now explore the model dependence of the non-charge-exchange isovector monopole  polarizability ($\alpha_M$). From our previous discussion and guidance from the dielectric theorem applied to a simple macroscopic model, we would expect for a given nucleus a linear relation of $\alpha_M J$ with the slope of the symmetry energy at saturation density $L$, as happens for the dipole polarizability \cite{Roca-Maza2018, Roca-Maza2013b, Roca-Maza2015}, provided the factor $\langle r^4\rangle-\langle r^2\rangle^2$ do not show a strong model dependence. This is actually the case as it is seen in Fig.~\ref{fig6}. All Skyrme EDF follow a good linear trend impliying the physics behind this quantiy is essentially dominated by the symmetry energy in commonly used EDFs. The dependence on the effective mass or incompressibilities is of much less relevance as it can be inferred by comparing the SAMi-J, SAMi-K and SAMi-m$^*$ families of systematically varied interactions \cite{Roca-Maza2013}.



\subsection{${}^{90}$Zr and ${}^{48}$Ca}
For completeness we show now the equivalent of Figs.~\ref{fig5} and \ref{fig6} for ${}^{90}$Zr and ${}^{48}$Ca. For the former nucleus, the conclusion is analogous to that of ${}^{208}$Pb: the centroid energies in the three channels do not depend much on the EDF employed and the centroid energy of the non-charge exchange channel follows the approximate rule given in Eq.~(\ref{const_app}). The monopole polarizability shows a clear correlation with the symmetry energy parameters although it is sizably degraded with respect to the heavier nucleus ${}^{208}$Pb, as it could naturally be expected. In ${}^{48}$Ca the situation remains qualitatively the same for the monopole polarizability but clearly changes for the case of the centroid energies. The centroid energies in this lighter nucleus are now much closer one to the other (cf. Fig.~\ref{fig2}). This magnifies the differences of centroid energies and highlights the expected inaccuracy of the approximate Eq.~(\ref{const_app}) for medium (and light) nuclei.  

\begin{figure*}[t!]
  \includegraphics[width=0.45\linewidth]{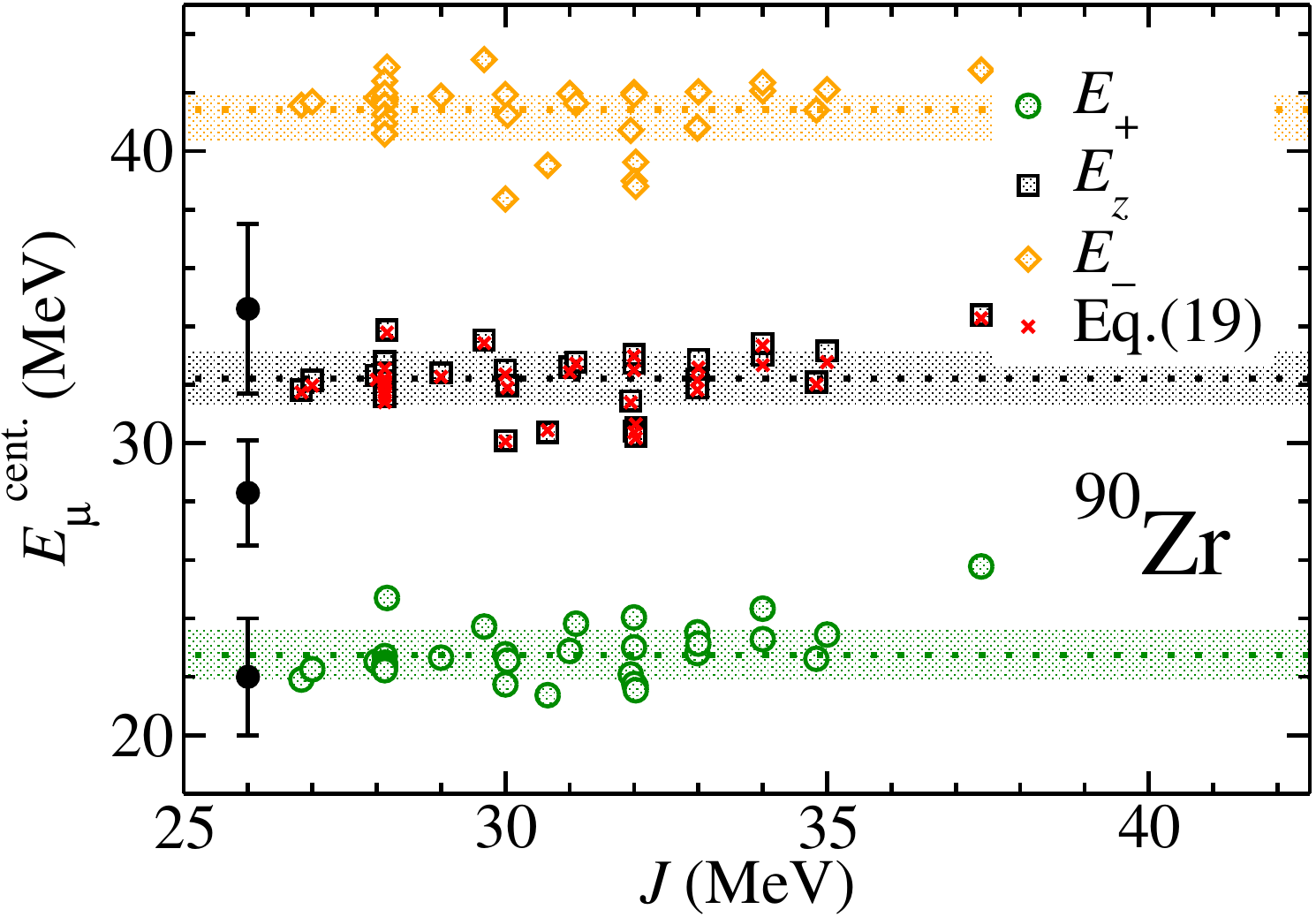}
  \hfill
  \includegraphics[width=0.45\linewidth]{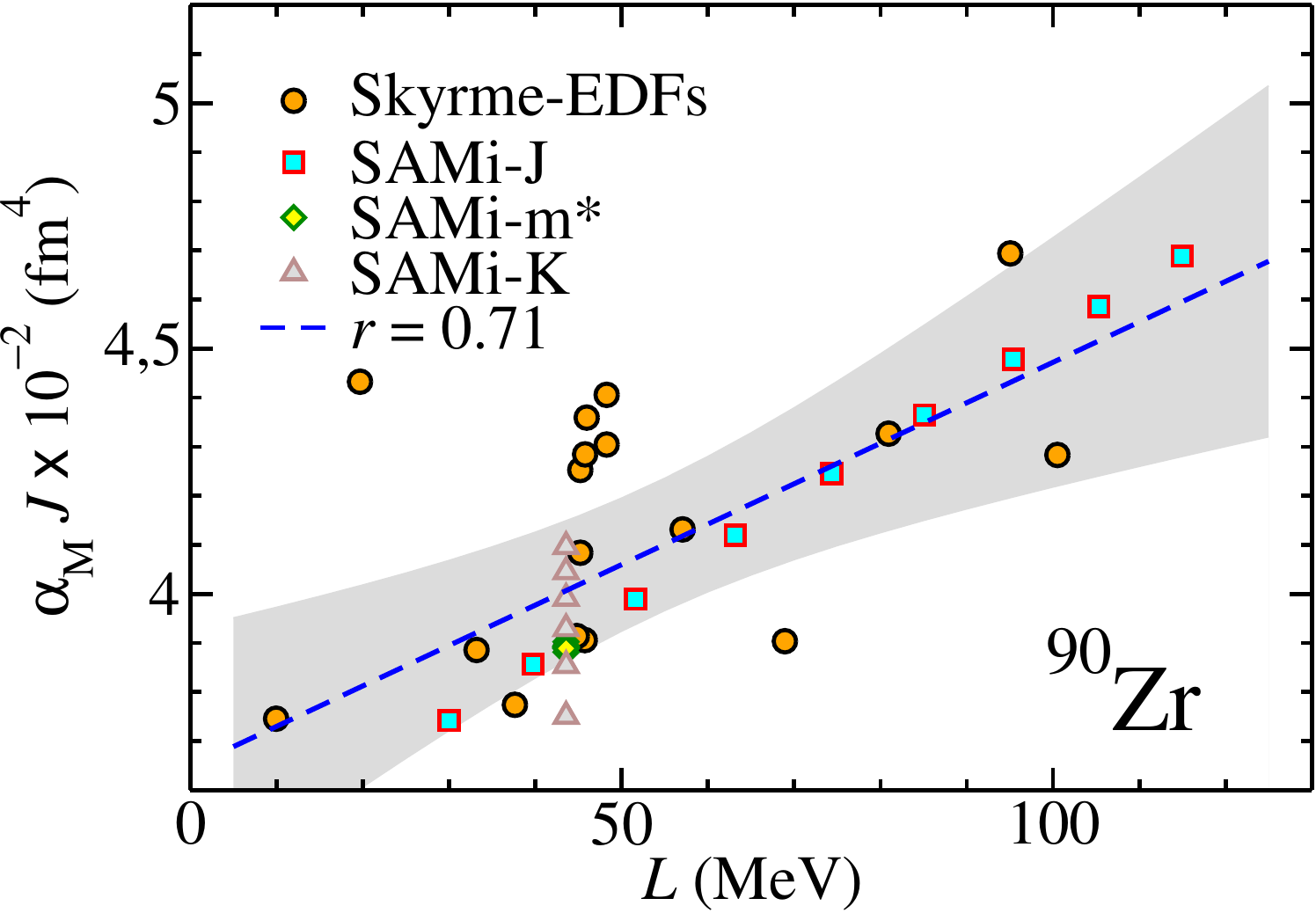}
  \caption{Left panel:  
  centroid excitation energy of the IVGMR in the three isospin chanels  as a function of the symmetry energy at saturation density. Data has been taken from \cite{Ere86}. Right panel: 
  non-charge-exchange isovector monopole  polarizability ($\alpha_M$) times the symmetry energy at saturation density as a function of the slope of the symmetery energy at saturation density. Results have been predicted by a set of 40 Skyrme EDFs. See the text for details. 
  \label{fig8}}
\end{figure*}

\begin{figure*}[t!]
  \includegraphics[width=0.45\linewidth]{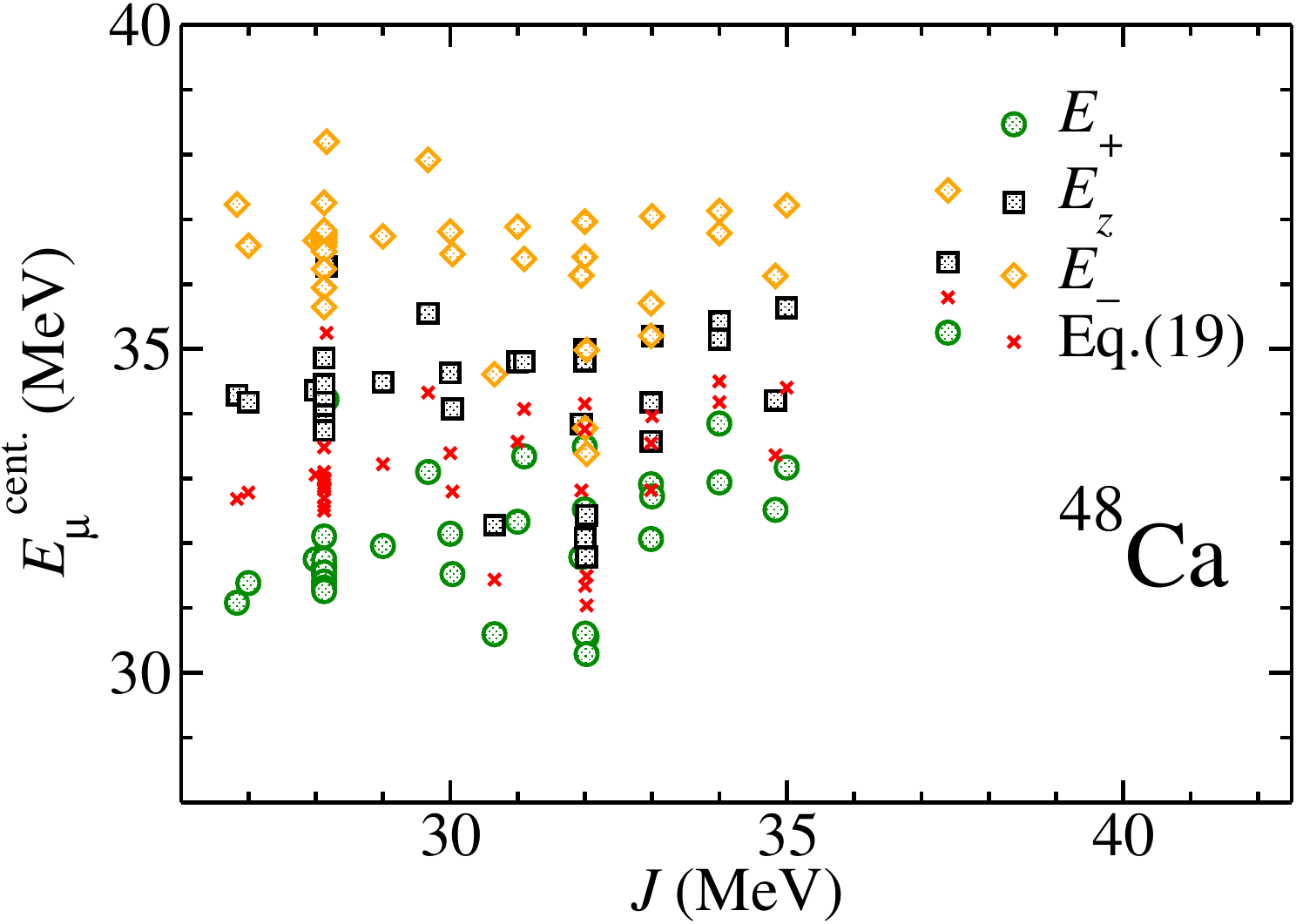}
  \hfill
  \includegraphics[width=0.45\linewidth]{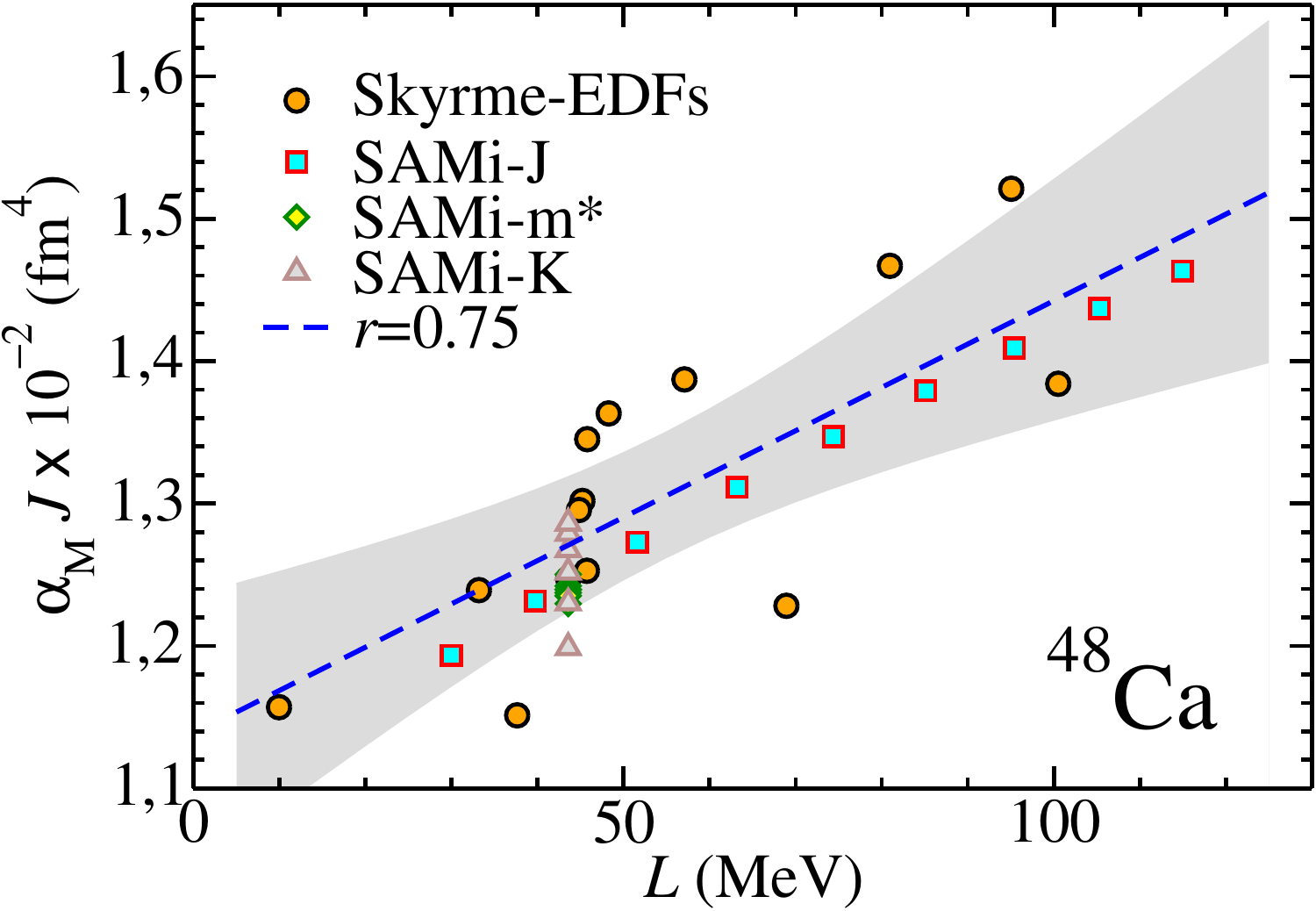}
  \caption{Left panel:  
  centroid excitation energy of the IVGMR in the three isospin chanels  as a function of the symmetry energy at saturation density. Data has been taken from \cite{Ere86}. Right panel: 
  non-charge-exchange isovector monopole  polarizability ($\alpha_M$) times the symmetry energy at saturation density as a function of the slope of the symmetery energy at saturation density. Results have been predicted by a set of 40 Skyrme EDFs. See the text for details. \label{fig9}}
\end{figure*}

\section{Conclusions}
\label{conclusions}

We have presented a systematic theoretical study of the three isospin channels of the isovector giant monopole resonance within self-consistent Skyrme Hartree-Fock plus Random-Phase Approximation. Particular attention has been devoted to defining excitation operators that isolate the genuine IVGMR response. In the charge-exchange channels, the standard monopole operator contains a large isobaric-analog-state contribution. We have shown that the modified operator \(\widetilde O_\pm\), obtained by subtracting the IAS 
contribution (estimated by 
knowing its
transition density) removes 
indeed the IAS while leaving the IVGMR region essentially unaltered. For the non-charge-exchange channel, the one-body center-of-mass correction is suppressed by inverse powers of the mass number and has a negligible effect for the nuclei considered here.

We have derived the NEWSR and EWSR associated with the modified charge-exchange operators and established relations between the charge-exchange and non-charge-exchange responses. In particular, the average of the corrected charge-exchange NEWSRs is found to be close to the non-charge-exchange NEWSR. This approximate relation emerges only after IAS subtraction. 

The centroid energies of the three IVGMR components display an almost symmetric pattern. The non-charge-exchange centroid is accurately reproduced by the average of the two charge-exchange centroids for \(^{90}\mathrm{Zr}\), and \(^{208}\mathrm{Pb}\). This result is understood in terms of the isospin decomposition of a collective excitation carrying one unit of isospin. The leading splitting is governed by the Coulomb displacement energy, the neutron-proton mass difference, and the finite-nucleus symmetry potential, whereas the isotensor contribution produces only a small departure from the symmetric pattern. The corresponding simple estimate provides a natural interpretation of the calculated splittings and of their weak dependence on the employed energy-density functional.

Finally, calculations based on a broad set of Skyrme EDFs show that, for a fixed nucleus, the centroid energies of the three channels vary only weakly with the symmetry energy, effective mass, and incompressibility. In contrast, the non-charge-exchange isovector monopole polarizability exhibits a pronounced correlation with the density dependence of the symmetry energy. The IVGMR polarizability may therefore provide complementary information on the isovector sector of nuclear EDFs, while the robust relation between the three centroid energies supports the extraction of the non-charge-exchange IVGMR energy from charge-exchange measurements in medium-heavy and heavy nuclei. Future measurements with improved radioactive-beam intensities and \(\gamma\)-tagged charge-exchange reactions should make this program experimentally accessible over a broader range of nuclei.

\begin{acknowledgments}
X.~R.~M. acknowledges support by
  MI-CIU/AEI/10.13039/501100011033 
  and by FEDER UE through grants PID2023-147112NB-C22 and CNS2025-165430;
  and through the ``Unit of Excellence Maria de Maeztu 2025-2028''
  award to the Institute of Cosmos Sciences, grant CEX2024-001451-M.
  Additional support is provided by the Generalitat de Catalunya (AGAUR) through grant 2021SGR01095.
\end{acknowledgments}

\appendix
\section{Center-of-mass correction} 
\label{app2}
Here we derive the one-body center of mass correction to the electromagnetic and isovector non-charge exchange monopole operators. For that we calculate the center-of-mass (CM) vector assuming equal masses of neutrons and protons giving
\begin{equation}
\vec{R}_{\rm CM} = \frac{1}{A}\sum_{i=1}^A \vec{r}_i \ . 
\end{equation}
The monopole electromagnetic (EM) operator including CM corrections will then be written as 
\begin{eqnarray}
\mathcal{O}_{\rm EM}^{\rm CM} &=& \sum_{i=1}^Z \left(\vec{r}_i - \vec{R}_{\rm CM}\right)^2 \nonumber\\
&=& \sum_{i=1}^Z r_i^2 + \sum_{i=1}^Z R_{\rm CM}^2  -2 \vec{R}_{\rm CM}\cdot\sum_{i=1}^Z\vec{r}_i \nonumber\\ 
&=& \sum_{i=1}^Z r_i^2 + \frac{Z}{A^2}\sum_{i,j=1}^A \vec{r}_i\cdot\vec{r_j}-\frac{2}{A}\sum_{i=1}^A \vec{r}_i\cdot\sum_{i=1}^Z\vec{r}_i \nonumber\\
\end{eqnarray}
which keeping only one-body terms results in 
\begin{eqnarray}
\mathcal{O}_{\rm EM}^{\rm CM} &=& \frac{A(A-1)}{A^2}\sum_{i=1}^Z r_i^2 + \frac{Z}{A^2}\sum_{i=1}^N r_i^2 - \frac{N}{A^2}\sum_{i=1}^Z r_i^2\nonumber\\
&=& \frac{A(A-1)}{A^2}\mathcal{O}_{\rm EM} + \frac{Z}{A^2}\sum_{i=1}^N r_i^2 - \frac{N}{A^2}\sum_{i=1}^Z r_i^2\nonumber\\
\end{eqnarray}
which is a correction of order $A^{-1}$ to the non CM corrected EM operator. The latter being re-scaled by a factor that is very close to one except for very light nuclei. 

\begin{figure}[t!]
  \includegraphics[width=\linewidth]{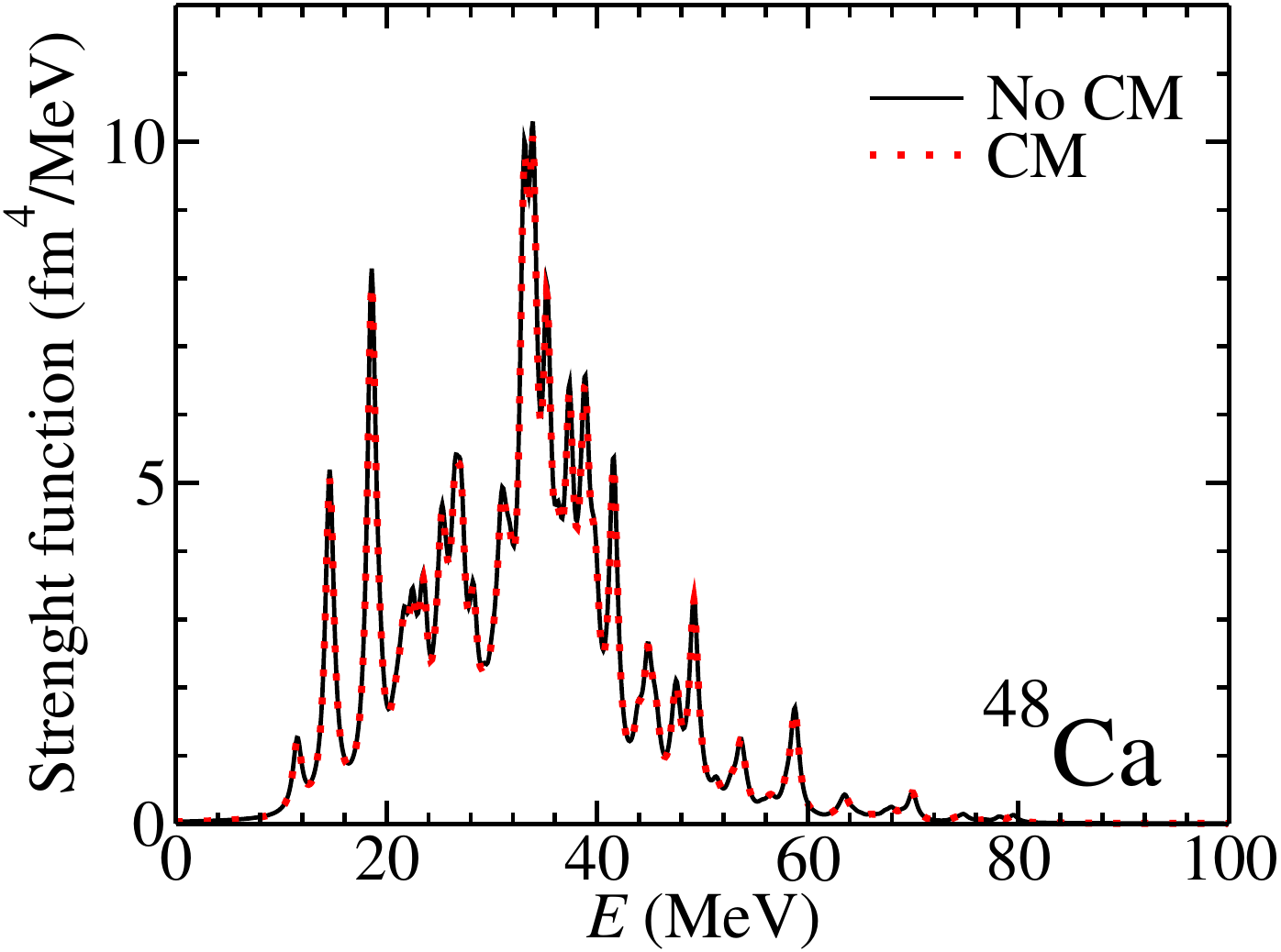}
  \caption{Non-charge-exchange IVGMR in ${}^{48}$Ca using $\mathcal{O}_z$ (black solid line labeled as No CM) and $\mathcal{O}_z^{\rm CM}$ (red dotted line labeled as CM) as defined in the text and Appendix~\ref{app2}. Predictions shown for the SAMi EDF. \label{figB1}}
\end{figure}

Regarding the isovector operator, following the same approximation, one can write 
\begin{eqnarray}
\mathcal{O}_z^{\rm CM} &=& \frac{A(A-1)}{A^2}\sum_{i=1}^Z r_i^2t_z(i) + \frac{N}{A^2}\sum_{i=1}^Z r_i^2 - \frac{Z}{A^2}\sum_{i=1}^N r_i^2\nonumber\\
&=& \frac{A(A-1)}{A^2}\mathcal{O}_z + \frac{Z}{A^2}\sum_{i=1}^N r_i^2 - \frac{N}{A^2}\sum_{i=1}^Z r_i^2 \ . \nonumber\\
\end{eqnarray}
One notes that the corection to both operators is the same but with opposite sign. That is, if we define
\begin{equation}
\mathcal{O}_{\rm corr}\equiv\frac{Z}{A}\sum_{i=1}^N r_i^2 - \frac{N}{A}\sum_{i=1}^Z r_i^2
\end{equation} 
analogous to the dipole operator, we can write, 
\begin{eqnarray}
\mathcal{O}_{\rm EM}^{\rm CM} &=& \left(1-\frac{1}{A}\right)\mathcal{O}_{\rm EM} + \frac{1}{A}\mathcal{O}_{\rm corr}\\
\mathcal{O}_z^{\rm CM} &=& \left(1-\frac{1}{A}\right)\mathcal{O}_z - \frac{1}{A}\mathcal{O}_{\rm corr}
\end{eqnarray}
Hence the CM correction to $\mathcal{O}_z$ or $\mathcal{O}_{\rm EM}$ is suppressed by two $1/A$ terms. In Fig.~\ref{figB1}, we show the comparison between $\mathcal{O}_z$
and $\mathcal{O}_z^{\rm CM}$ for the lightest nucleus studied here where the effects should be larger. The effect of the CM is negligible, about 2\% reduction in the strength as well as in the NEWSR and EWSR and below 0.05\% in the centroid energy. For this test we used the SAMi EDF. 

\section{Double communtator for the new IVM operator $\tilde{\mathcal{O}}_\pm$}
\label{app1}
Here we evaluate the expectation value of the double commutator for the IVM operator $\tilde{\mathcal{O}}_\pm$ proposed in this work. Starting from the left hand side of Eq.~(\ref{dc}) and writing $A_\pm\equiv\mathcal{O}_\pm$ and $B_\pm\equiv\langle r^2_{\rm exc}\rangle\mathcal{O}_\pm^{\rm IAS}$ to make the notation lighter,  
\begin{eqnarray}  
&&  \langle 0 \vert[\tilde{\mathcal{O}}_+,[\mathcal{H},\tilde{\mathcal{O}}_-]] \vert 0\rangle =\langle 0 \vert[A_++B_+,[\mathcal{H},A_-+B_-]] \vert 0\rangle\nonumber\\
  &&~~~~~~=\langle 0\vert [A_+,[\mathcal{H},A_-]]\vert 0\rangle
  +\langle 0\vert [B_+,[\mathcal{H},B_-]]\vert 0\rangle\nonumber\\
  &&~~~~~~-\langle 0\vert [B_+,[\mathcal{H},A_-]]\vert 0\rangle
  -\langle 0\vert [A_+,[\mathcal{H},B_-]]\vert 0\rangle
\end{eqnarray}
according to the identity $[B_+,[\mathcal{H},A_-]]+[\mathcal{H},[A_-,B_+]]+[A_-,[B_+,\mathcal{H}]]=0$, one may write
\begin{eqnarray}  
&&  \langle 0 \vert[\tilde{\mathcal{O}}_+,[\mathcal{H},\tilde{\mathcal{O}}_-]] \vert 0\rangle =\langle 0\vert [A_+,[\mathcal{H},A_-]]\vert 0\rangle\nonumber\\
  &&~~~~~~+\langle 0\vert [B_+,[\mathcal{H},B_-]]\vert 0\rangle
  -\langle 0\vert [A_-,[\mathcal{H},B_+]]\vert 0\rangle\nonumber\\
  &&~~~~~~-\langle 0\vert [A_+,[\mathcal{H},B_-]]\vert 0\rangle
  -\langle 0\vert [\mathcal{H},[A_-,B_+]]\vert 0\rangle\nonumber\\
\end{eqnarray}
since $\mathcal{H}$ commutes with $\sum_i\tau_z(i)$, the last term in r.h.s. do not contribute. It is important to note that for our specific choice of the operators
\begin{equation}
  [A_-,[\mathcal{H},B_+]]^\dag=[A_+,[\mathcal{H},B_-]]
\end{equation}
and, thus, both will give the same expectation value. We can write then,
\begin{eqnarray}  
\langle 0 \vert[\tilde{\mathcal{O}}_+,[\mathcal{H},\tilde{\mathcal{O}}_-]] \vert 0\rangle &=&\langle 0\vert [A_+,[\mathcal{H},A_-]]\vert 0\rangle\nonumber\\
  &+&\langle 0\vert [B_+-2A_+,[\mathcal{H},B_-]]\vert 0\rangle\nonumber\\
\end{eqnarray}
which comming back to the original notation,
\begin{eqnarray}  
&&\langle 0 \vert[\tilde{\mathcal{O}}_+,[\mathcal{H},\tilde{\mathcal{O}}_-]] \vert 0\rangle =\langle 0\vert [\mathcal{O}_+,[\mathcal{H},\mathcal{O}_-]]\vert 0\rangle\nonumber\\
  &&~~~~+\frac{1}{4\pi}\langle 0\vert [\sum_{ij}\left(\langle r_{\rm exc}^2\rangle^2-2r_i^2\langle r_{\rm exc}^2\rangle\right)t_+(i),[\mathcal{H},t_-(j)]]\vert 0\rangle\nonumber\\
\end{eqnarray}
Since  $[\mathcal{H},t_-]$ selects only isospin symmetry breaking terms in the Hamiltonian, that is, essentially Coulomb, one could simplify the previous expression as 
\begin{eqnarray}  
&&\langle 0 \vert[\tilde{\mathcal{O}}_+,[\mathcal{H},\tilde{\mathcal{O}}_-]] \vert 0\rangle =\langle 0\vert [\mathcal{O}_+,[\mathcal{H},\mathcal{O}_-]]\vert 0\rangle\nonumber\\
  &&~~~~+\frac{1}{4\pi}\langle 0\vert\sum_{ij}\left(\langle r_{\rm exc}^2\rangle^2-2r_i^2\langle r_{\rm exc}^2\rangle\right) [t_+(i),[\mathcal{H},t_-(j)]]\vert 0\rangle\nonumber\\
\end{eqnarray}
Now considering that the double commutator defining the $\eta_C$ term in Ref.~\cite{Auerbach83} when using the original IVM operator $\mathcal{O}_\pm$ is
\begin{eqnarray}  
\frac{1}{4\pi}\langle 0\vert \sum_{ij}r_i^2r_j^2[t_+(i),[\mathcal{H},t_-(j)]]\vert 0\rangle
\end{eqnarray}
one can just redefine it when using the new operator $\tilde{\mathcal{O}}_\pm$ as 
\begin{eqnarray}  
  \frac{1}{4\pi}\langle 0\vert\sum_{ij}(r_i^2-\langle r_{\rm exc}^2\rangle)(r_j^2-\langle r_{\rm exc}^2\rangle) [t_+(i),[\mathcal{H},t_-(j)]]\vert 0\rangle\nonumber\\
\end{eqnarray}
This is the expression used in Eq.~(\ref{eta_c}).

\bibliography{bibliography}

\begin{thebibliography}{49}%
\makeatletter
\providecommand \@ifxundefined [1]{%
 \@ifx{#1\undefined}
}%
\providecommand \@ifnum [1]{%
 \ifnum #1\expandafter \@firstoftwo
 \else \expandafter \@secondoftwo
 \fi
}%
\providecommand \@ifx [1]{%
 \ifx #1\expandafter \@firstoftwo
 \else \expandafter \@secondoftwo
 \fi
}%
\providecommand \natexlab [1]{#1}%
\providecommand \enquote  [1]{``#1''}%
\providecommand \bibnamefont  [1]{#1}%
\providecommand \bibfnamefont [1]{#1}%
\providecommand \citenamefont [1]{#1}%
\providecommand \href@noop [0]{\@secondoftwo}%
\providecommand \href [0]{\begingroup \@sanitize@url \@href}%
\providecommand \@href[1]{\@@startlink{#1}\@@href}%
\providecommand \@@href[1]{\endgroup#1\@@endlink}%
\providecommand \@sanitize@url [0]{\catcode `\\12\catcode `\$12\catcode
  `\&12\catcode `\#12\catcode `\^12\catcode `\_12\catcode `\%12\relax}%
\providecommand \@@startlink[1]{}%
\providecommand \@@endlink[0]{}%
\providecommand \url  [0]{\begingroup\@sanitize@url \@url }%
\providecommand \@url [1]{\endgroup\@href {#1}{\urlprefix }}%
\providecommand \urlprefix  [0]{URL }%
\providecommand \Eprint [0]{\href }%
\providecommand \doibase [0]{http://dx.doi.org/}%
\providecommand \selectlanguage [0]{\@gobble}%
\providecommand \bibinfo  [0]{\@secondoftwo}%
\providecommand \bibfield  [0]{\@secondoftwo}%
\providecommand \translation [1]{[#1]}%
\providecommand \BibitemOpen [0]{}%
\providecommand \bibitemStop [0]{}%
\providecommand \bibitemNoStop [0]{.\EOS\space}%
\providecommand \EOS [0]{\spacefactor3000\relax}%
\providecommand \BibitemShut  [1]{\csname bibitem#1\endcsname}%
\let\auto@bib@innerbib\@empty
\bibitem [{\citenamefont {Roca-Maza}\ and\ \citenamefont
  {Paar}(2018)}]{Roca-Maza2018}%
  \BibitemOpen
  \bibfield  {author} {\bibinfo {author} {\bibfnamefont {X.}~\bibnamefont
  {Roca-Maza}}\ and\ \bibinfo {author} {\bibfnamefont {N.}~\bibnamefont
  {Paar}},\ }\bibfield  {title} {\enquote {\bibinfo {title} {Nuclear equation
  of state from ground and collective excited state properties of nuclei},}\
  }\href {\doibase 10.1016/j.ppnp.2018.04.001} {\bibfield  {journal} {\bibinfo
  {journal} {Prog. Part. Nucl. Phys.}\ }\textbf {\bibinfo {volume} {101}},\
  \bibinfo {pages} {96--176} (\bibinfo {year} {2018})}\BibitemShut {NoStop}%
\bibitem [{\citenamefont {Tanihata}\ \emph {et~al.}(2020)\citenamefont
  {Tanihata}, \citenamefont {Toki},\ and\ \citenamefont {Kajino}}]{hnp}%
  \BibitemOpen
  \bibinfo {editor} {\bibfnamefont {Isao}\ \bibnamefont {Tanihata}}, \bibinfo
  {editor} {\bibfnamefont {Hiroshi}\ \bibnamefont {Toki}}, \ and\ \bibinfo
  {editor} {\bibfnamefont {Toshitaka}\ \bibnamefont {Kajino}},\ eds.,\ in\
  \href {\doibase 10.1007/978-981-19-6345-2} {\emph {\bibinfo {booktitle}
  {Handbook of Nuclear Physics}}}\ (\bibinfo  {publisher} {Springer Nature
  Singapore},\ \bibinfo {address} {Singapore},\ \bibinfo {year}
  {2020})\BibitemShut {NoStop}%
\bibitem [{\citenamefont {Garg}\ and\ \citenamefont {Colò}(2018)}]{Gar2018}%
  \BibitemOpen
  \bibfield  {author} {\bibinfo {author} {\bibfnamefont {Umesh}\ \bibnamefont
  {Garg}}\ and\ \bibinfo {author} {\bibfnamefont {Gianluca}\ \bibnamefont
  {Colò}},\ }\bibfield  {title} {\enquote {\bibinfo {title} {The
  compression-mode giant resonances and nuclear incompressibility},}\ }\href
  {\doibase https://doi.org/10.1016/j.ppnp.2018.03.001} {\bibfield  {journal}
  {\bibinfo  {journal} {Progress in Particle and Nuclear Physics}\ }\textbf
  {\bibinfo {volume} {101}},\ \bibinfo {pages} {55--95} (\bibinfo {year}
  {2018})}\BibitemShut {NoStop}%
\bibitem [{\citenamefont {Zegers}(2020)}]{Zeg2020}%
  \BibitemOpen
  \bibfield  {author} {\bibinfo {author} {\bibfnamefont {Remco G.~T.}\
  \bibnamefont {Zegers}},\ }\enquote {\bibinfo {title} {Excitation of isovector
  giant resonances through charge-exchange reactions},}\ in\ \href {\doibase
  10.1007/978-981-15-8818-1_77-1} {\emph {\bibinfo {booktitle} {Handbook of
  Nuclear Physics}}},\ \bibinfo {editor} {edited by\ \bibinfo {editor}
  {\bibfnamefont {Isao}\ \bibnamefont {Tanihata}}, \bibinfo {editor}
  {\bibfnamefont {Hiroshi}\ \bibnamefont {Toki}}, \ and\ \bibinfo {editor}
  {\bibfnamefont {Toshitaka}\ \bibnamefont {Kajino}}}\ (\bibinfo  {publisher}
  {Springer Nature Singapore},\ \bibinfo {address} {Singapore},\ \bibinfo
  {year} {2020})\ pp.\ \bibinfo {pages} {1--35}\BibitemShut {NoStop}%
\bibitem [{\citenamefont {Bowman}\ \emph {et~al.}(1987)\citenamefont {Bowman},
  \citenamefont {Lipparini},\ and\ \citenamefont {Stringari}}]{bowman1987}%
  \BibitemOpen
  \bibfield  {author} {\bibinfo {author} {\bibfnamefont {J.D.}\ \bibnamefont
  {Bowman}}, \bibinfo {author} {\bibfnamefont {E.}~\bibnamefont {Lipparini}}, \
  and\ \bibinfo {author} {\bibfnamefont {S.}~\bibnamefont {Stringari}},\
  }\bibfield  {title} {\enquote {\bibinfo {title} {Isovector monopole
  excitation energies},}\ }\href {\doibase
  https://doi.org/10.1016/0370-2693(87)91041-0} {\bibfield  {journal} {\bibinfo
   {journal} {Physics Letters B}\ }\textbf {\bibinfo {volume} {197}},\ \bibinfo
  {pages} {497--499} (\bibinfo {year} {1987})}\BibitemShut {NoStop}%
\bibitem [{\citenamefont {Oertel}\ \emph {et~al.}(2017)\citenamefont {Oertel},
  \citenamefont {Hempel}, \citenamefont {Kl{\"a}hn},\ and\ \citenamefont
  {Typel}}]{Oertel2017}%
  \BibitemOpen
  \bibfield  {author} {\bibinfo {author} {\bibfnamefont {M.}~\bibnamefont
  {Oertel}}, \bibinfo {author} {\bibfnamefont {M.}~\bibnamefont {Hempel}},
  \bibinfo {author} {\bibfnamefont {T.}~\bibnamefont {Kl{\"a}hn}}, \ and\
  \bibinfo {author} {\bibfnamefont {S.}~\bibnamefont {Typel}},\ }\bibfield
  {title} {\enquote {\bibinfo {title} {Equations of state for supernovae and
  compact stars},}\ }\href {\doibase 10.1103/RevModPhys.89.015007} {\bibfield
  {journal} {\bibinfo  {journal} {Rev. Mod. Phys.}\ }\textbf {\bibinfo {volume}
  {89}},\ \bibinfo {pages} {015007} (\bibinfo {year} {2017})}\BibitemShut
  {NoStop}%
\bibitem [{\citenamefont {{\it et al.}}(2024)}]{Sorensen2024}%
  \BibitemOpen
  \bibfield  {author} {\bibinfo {author} {\bibfnamefont {Agnieszka~Sorensen}\
  \bibnamefont {{\it et al.}}},\ }\bibfield  {title} {\enquote {\bibinfo
  {title} {Dense nuclear matter equation of state from heavy-ion collisions},}\
  }\href {\doibase https://doi.org/10.1016/j.ppnp.2023.104080} {\bibfield
  {journal} {\bibinfo  {journal} {Progress in Particle and Nuclear Physics}\
  }\textbf {\bibinfo {volume} {134}},\ \bibinfo {pages} {104080} (\bibinfo
  {year} {2024})}\BibitemShut {NoStop}%
\bibitem [{\citenamefont {Erell}\ \emph {et~al.}(1984)\citenamefont {Erell},
  \citenamefont {Alster}, \citenamefont {Lichtenstadt}, \citenamefont
  {Moinester}, \citenamefont {Bowman}, \citenamefont {Cooper}, \citenamefont
  {Irom}, \citenamefont {Matis}, \citenamefont {Piasetzky}, \citenamefont
  {Sennhauser},\ and\ \citenamefont {Ingram}}]{Erell84}%
  \BibitemOpen
  \bibfield  {author} {\bibinfo {author} {\bibfnamefont {A.}~\bibnamefont
  {Erell}}, \bibinfo {author} {\bibfnamefont {J.}~\bibnamefont {Alster}},
  \bibinfo {author} {\bibfnamefont {J.}~\bibnamefont {Lichtenstadt}}, \bibinfo
  {author} {\bibfnamefont {M.~A.}\ \bibnamefont {Moinester}}, \bibinfo {author}
  {\bibfnamefont {J.~D.}\ \bibnamefont {Bowman}}, \bibinfo {author}
  {\bibfnamefont {M.~D.}\ \bibnamefont {Cooper}}, \bibinfo {author}
  {\bibfnamefont {F.}~\bibnamefont {Irom}}, \bibinfo {author} {\bibfnamefont
  {H.~S.}\ \bibnamefont {Matis}}, \bibinfo {author} {\bibfnamefont
  {E.}~\bibnamefont {Piasetzky}}, \bibinfo {author} {\bibfnamefont
  {U.}~\bibnamefont {Sennhauser}}, \ and\ \bibinfo {author} {\bibfnamefont
  {Q.}~\bibnamefont {Ingram}},\ }\bibfield  {title} {\enquote {\bibinfo {title}
  {Properties of the isovector monopole and other giant resonances in pion
  charge exchange},}\ }\href {\doibase 10.1103/PhysRevLett.52.2134} {\bibfield
  {journal} {\bibinfo  {journal} {Phys. Rev. Lett.}\ }\textbf {\bibinfo
  {volume} {52}},\ \bibinfo {pages} {2134--2137} (\bibinfo {year}
  {1984})}\BibitemShut {NoStop}%
\bibitem [{\citenamefont {Erell}\ \emph {et~al.}(1986)\citenamefont {Erell},
  \citenamefont {Alster}, \citenamefont {Lichtenstadt}, \citenamefont
  {Moinester}, \citenamefont {Bowman}, \citenamefont {Cooper}, \citenamefont
  {Irom}, \citenamefont {Matis}, \citenamefont {Piasetzky},\ and\ \citenamefont
  {Sennhauser}}]{Ere86}%
  \BibitemOpen
  \bibfield  {author} {\bibinfo {author} {\bibfnamefont {A.}~\bibnamefont
  {Erell}}, \bibinfo {author} {\bibfnamefont {J.}~\bibnamefont {Alster}},
  \bibinfo {author} {\bibfnamefont {J.}~\bibnamefont {Lichtenstadt}}, \bibinfo
  {author} {\bibfnamefont {M.~A.}\ \bibnamefont {Moinester}}, \bibinfo {author}
  {\bibfnamefont {J.~D.}\ \bibnamefont {Bowman}}, \bibinfo {author}
  {\bibfnamefont {M.~D.}\ \bibnamefont {Cooper}}, \bibinfo {author}
  {\bibfnamefont {F.}~\bibnamefont {Irom}}, \bibinfo {author} {\bibfnamefont
  {H.~S.}\ \bibnamefont {Matis}}, \bibinfo {author} {\bibfnamefont
  {E.}~\bibnamefont {Piasetzky}}, \ and\ \bibinfo {author} {\bibfnamefont
  {U.}~\bibnamefont {Sennhauser}},\ }\bibfield  {title} {\enquote {\bibinfo
  {title} {Measurements on isovector giant resonances in pion charge
  exchange},}\ }\href {\doibase 10.1103/PhysRevC.34.1822} {\bibfield  {journal}
  {\bibinfo  {journal} {Phys. Rev. C}\ }\textbf {\bibinfo {volume} {34}},\
  \bibinfo {pages} {1822--1844} (\bibinfo {year} {1986})}\BibitemShut {NoStop}%
\bibitem [{\citenamefont {Irom}\ \emph {et~al.}(1986)\citenamefont {Irom},
  \citenamefont {Bowman}, \citenamefont {Bolme}, \citenamefont {Piasetzky},
  \citenamefont {Sennhauser}, \citenamefont {Alster}, \citenamefont
  {Lichtenstadt}, \citenamefont {Moinester}, \citenamefont {Knudson},
  \citenamefont {Rokni},\ and\ \citenamefont {Siciliano}}]{Irom86}%
  \BibitemOpen
  \bibfield  {author} {\bibinfo {author} {\bibfnamefont {F.}~\bibnamefont
  {Irom}}, \bibinfo {author} {\bibfnamefont {J.~D.}\ \bibnamefont {Bowman}},
  \bibinfo {author} {\bibfnamefont {G.~O.}\ \bibnamefont {Bolme}}, \bibinfo
  {author} {\bibfnamefont {E.}~\bibnamefont {Piasetzky}}, \bibinfo {author}
  {\bibfnamefont {U.}~\bibnamefont {Sennhauser}}, \bibinfo {author}
  {\bibfnamefont {J.}~\bibnamefont {Alster}}, \bibinfo {author} {\bibfnamefont
  {J.}~\bibnamefont {Lichtenstadt}}, \bibinfo {author} {\bibfnamefont
  {M.}~\bibnamefont {Moinester}}, \bibinfo {author} {\bibfnamefont {J.~N.}\
  \bibnamefont {Knudson}}, \bibinfo {author} {\bibfnamefont {S.~H.}\
  \bibnamefont {Rokni}}, \ and\ \bibinfo {author} {\bibfnamefont {E.~R.}\
  \bibnamefont {Siciliano}},\ }\bibfield  {title} {\enquote {\bibinfo {title}
  {Excitation of isovector giant resonances in pion single-charge exchange at
  120, 165, and 230 mev},}\ }\href {\doibase 10.1103/PhysRevC.34.2231}
  {\bibfield  {journal} {\bibinfo  {journal} {Phys. Rev. C}\ }\textbf {\bibinfo
  {volume} {34}},\ \bibinfo {pages} {2231--2239} (\bibinfo {year}
  {1986})}\BibitemShut {NoStop}%
\bibitem [{\citenamefont {Nakayama}\ \emph {et~al.}(1999)\citenamefont
  {Nakayama}, \citenamefont {Akimune}, \citenamefont {Arimoto}, \citenamefont
  {Daito}, \citenamefont {Fujimura}, \citenamefont {Fujita}, \citenamefont
  {Fujiwara}, \citenamefont {Fushimi}, \citenamefont {Kohri}, \citenamefont
  {Koori}, \citenamefont {Takahisa}, \citenamefont {Takeuchi}, \citenamefont
  {Tamii}, \citenamefont {Tanaka}, \citenamefont {Yamagata}, \citenamefont
  {Yamamoto}, \citenamefont {Yonehara},\ and\ \citenamefont
  {Yoshida}}]{Nakayama1999}%
  \BibitemOpen
  \bibfield  {author} {\bibinfo {author} {\bibfnamefont {S.}~\bibnamefont
  {Nakayama}}, \bibinfo {author} {\bibfnamefont {H.}~\bibnamefont {Akimune}},
  \bibinfo {author} {\bibfnamefont {Y.}~\bibnamefont {Arimoto}}, \bibinfo
  {author} {\bibfnamefont {I.}~\bibnamefont {Daito}}, \bibinfo {author}
  {\bibfnamefont {H.}~\bibnamefont {Fujimura}}, \bibinfo {author}
  {\bibfnamefont {Y.}~\bibnamefont {Fujita}}, \bibinfo {author} {\bibfnamefont
  {M.}~\bibnamefont {Fujiwara}}, \bibinfo {author} {\bibfnamefont
  {K.}~\bibnamefont {Fushimi}}, \bibinfo {author} {\bibfnamefont
  {H.}~\bibnamefont {Kohri}}, \bibinfo {author} {\bibfnamefont
  {N.}~\bibnamefont {Koori}}, \bibinfo {author} {\bibfnamefont
  {K.}~\bibnamefont {Takahisa}}, \bibinfo {author} {\bibfnamefont
  {T.}~\bibnamefont {Takeuchi}}, \bibinfo {author} {\bibfnamefont
  {A.}~\bibnamefont {Tamii}}, \bibinfo {author} {\bibfnamefont
  {M.}~\bibnamefont {Tanaka}}, \bibinfo {author} {\bibfnamefont
  {T.}~\bibnamefont {Yamagata}}, \bibinfo {author} {\bibfnamefont
  {Y.}~\bibnamefont {Yamamoto}}, \bibinfo {author} {\bibfnamefont
  {K.}~\bibnamefont {Yonehara}}, \ and\ \bibinfo {author} {\bibfnamefont
  {H.}~\bibnamefont {Yoshida}},\ }\bibfield  {title} {\enquote {\bibinfo
  {title} {Isovector electric monopole resonance in ${}^{60}$ni},}\ }\href
  {\doibase 10.1103/PhysRevLett.83.690} {\bibfield  {journal} {\bibinfo
  {journal} {Phys. Rev. Lett.}\ }\textbf {\bibinfo {volume} {83}},\ \bibinfo
  {pages} {690--693} (\bibinfo {year} {1999})}\BibitemShut {NoStop}%
\bibitem [{\citenamefont {Scott}\ \emph {et~al.}(2017)\citenamefont {Scott},
  \citenamefont {Zegers}, \citenamefont {Almus}, \citenamefont {Austin},
  \citenamefont {Bazin}, \citenamefont {Brown}, \citenamefont {Campbell},
  \citenamefont {Gade}, \citenamefont {Bowry}, \citenamefont {Gal\`es},
  \citenamefont {Garg}, \citenamefont {Harakeh}, \citenamefont {Kwan},
  \citenamefont {Langer}, \citenamefont {Loelius}, \citenamefont {Lipschutz},
  \citenamefont {Litvinova}, \citenamefont {Lunderberg}, \citenamefont {Morse},
  \citenamefont {Noji}, \citenamefont {Perdikakis}, \citenamefont {Redpath},
  \citenamefont {Robin}, \citenamefont {Sakai}, \citenamefont {Sasamoto},
  \citenamefont {Sasano}, \citenamefont {Sullivan}, \citenamefont {Tostevin},
  \citenamefont {Uesaka},\ and\ \citenamefont {Weisshaar}}]{Sco2017}%
  \BibitemOpen
  \bibfield  {author} {\bibinfo {author} {\bibfnamefont {M.}~\bibnamefont
  {Scott}}, \bibinfo {author} {\bibfnamefont {R.~G.~T.}\ \bibnamefont
  {Zegers}}, \bibinfo {author} {\bibfnamefont {R.}~\bibnamefont {Almus}},
  \bibinfo {author} {\bibfnamefont {Sam~M.}\ \bibnamefont {Austin}}, \bibinfo
  {author} {\bibfnamefont {D.}~\bibnamefont {Bazin}}, \bibinfo {author}
  {\bibfnamefont {B.~A.}\ \bibnamefont {Brown}}, \bibinfo {author}
  {\bibfnamefont {C.}~\bibnamefont {Campbell}}, \bibinfo {author}
  {\bibfnamefont {A.}~\bibnamefont {Gade}}, \bibinfo {author} {\bibfnamefont
  {M.}~\bibnamefont {Bowry}}, \bibinfo {author} {\bibfnamefont
  {S.}~\bibnamefont {Gal\`es}}, \bibinfo {author} {\bibfnamefont
  {U.}~\bibnamefont {Garg}}, \bibinfo {author} {\bibfnamefont {M.~N.}\
  \bibnamefont {Harakeh}}, \bibinfo {author} {\bibfnamefont {E.}~\bibnamefont
  {Kwan}}, \bibinfo {author} {\bibfnamefont {C.}~\bibnamefont {Langer}},
  \bibinfo {author} {\bibfnamefont {C.}~\bibnamefont {Loelius}}, \bibinfo
  {author} {\bibfnamefont {S.}~\bibnamefont {Lipschutz}}, \bibinfo {author}
  {\bibfnamefont {E.}~\bibnamefont {Litvinova}}, \bibinfo {author}
  {\bibfnamefont {E.}~\bibnamefont {Lunderberg}}, \bibinfo {author}
  {\bibfnamefont {C.}~\bibnamefont {Morse}}, \bibinfo {author} {\bibfnamefont
  {S.}~\bibnamefont {Noji}}, \bibinfo {author} {\bibfnamefont {G.}~\bibnamefont
  {Perdikakis}}, \bibinfo {author} {\bibfnamefont {T.}~\bibnamefont {Redpath}},
  \bibinfo {author} {\bibfnamefont {C.}~\bibnamefont {Robin}}, \bibinfo
  {author} {\bibfnamefont {H.}~\bibnamefont {Sakai}}, \bibinfo {author}
  {\bibfnamefont {Y.}~\bibnamefont {Sasamoto}}, \bibinfo {author}
  {\bibfnamefont {M.}~\bibnamefont {Sasano}}, \bibinfo {author} {\bibfnamefont
  {C.}~\bibnamefont {Sullivan}}, \bibinfo {author} {\bibfnamefont {J.~A.}\
  \bibnamefont {Tostevin}}, \bibinfo {author} {\bibfnamefont {T.}~\bibnamefont
  {Uesaka}}, \ and\ \bibinfo {author} {\bibfnamefont {D.}~\bibnamefont
  {Weisshaar}},\ }\bibfield  {title} {\enquote {\bibinfo {title} {Observation
  of the isovector giant monopole resonance via the
  $^{28}\mathrm{Si}(^{10}\mathrm{Be},^{10}{\mathrm{b}}^{*}[1.74\text{ }\text{
  }\mathrm{MeV}])$ reaction at $100\text{ }\text{ }a\mathrm{MeV}$},}\ }\href
  {\doibase 10.1103/PhysRevLett.118.172501} {\bibfield  {journal} {\bibinfo
  {journal} {Phys. Rev. Lett.}\ }\textbf {\bibinfo {volume} {118}},\ \bibinfo
  {pages} {172501} (\bibinfo {year} {2017})}\BibitemShut {NoStop}%
\bibitem [{\citenamefont {Zegers}(2023)}]{Zegers2023}%
  \BibitemOpen
  \bibfield  {author} {\bibinfo {author} {\bibfnamefont {Remco G.~T.}\
  \bibnamefont {Zegers}},\ }\enquote {\bibinfo {title} {Excitation of isovector
  giant resonances through charge-exchange reactions},}\ in\ \href {\doibase
  10.1007/978-981-19-6345-2_77} {\emph {\bibinfo {booktitle} {Handbook of
  Nuclear Physics}}},\ \bibinfo {editor} {edited by\ \bibinfo {editor}
  {\bibfnamefont {Isao}\ \bibnamefont {Tanihata}}, \bibinfo {editor}
  {\bibfnamefont {Hiroshi}\ \bibnamefont {Toki}}, \ and\ \bibinfo {editor}
  {\bibfnamefont {Toshitaka}\ \bibnamefont {Kajino}}}\ (\bibinfo  {publisher}
  {Springer Nature Singapore},\ \bibinfo {address} {Singapore},\ \bibinfo
  {year} {2023})\ pp.\ \bibinfo {pages} {739--773}\BibitemShut {NoStop}%
\bibitem [{\citenamefont {Schunck}(2019)}]{sch2019}%
  \BibitemOpen
  \bibinfo {editor} {\bibfnamefont {Nicolas}\ \bibnamefont {Schunck}},\ ed.,\
  \href {\doibase 10.1088/2053-2563/aae0ed} {\emph {\bibinfo {title} {Energy
  Density Functional Methods for Atomic Nuclei}}},\ 2053-2563\ (\bibinfo
  {publisher} {IOP Publishing},\ \bibinfo {year} {2019})\BibitemShut {NoStop}%
\bibitem [{\citenamefont {Bender}\ \emph {et~al.}(2003)\citenamefont {Bender},
  \citenamefont {Heenen},\ and\ \citenamefont {Reinhard}}]{Ben2003}%
  \BibitemOpen
  \bibfield  {author} {\bibinfo {author} {\bibfnamefont {Michael}\ \bibnamefont
  {Bender}}, \bibinfo {author} {\bibfnamefont {Paul-Henri}\ \bibnamefont
  {Heenen}}, \ and\ \bibinfo {author} {\bibfnamefont {Paul-Gerhard}\
  \bibnamefont {Reinhard}},\ }\bibfield  {title} {\enquote {\bibinfo {title}
  {Self-consistent mean-field models for nuclear structure},}\ }\href {\doibase
  10.1103/RevModPhys.75.121} {\bibfield  {journal} {\bibinfo  {journal} {Rev.
  Mod. Phys.}\ }\textbf {\bibinfo {volume} {75}},\ \bibinfo {pages} {121--180}
  (\bibinfo {year} {2003})}\BibitemShut {NoStop}%
\bibitem [{\citenamefont {Paar}\ \emph {et~al.}(2007)\citenamefont {Paar},
  \citenamefont {Vretenar}, \citenamefont {Khan},\ and\ \citenamefont
  {Colò}}]{Paa2007}%
  \BibitemOpen
  \bibfield  {author} {\bibinfo {author} {\bibfnamefont {Nils}\ \bibnamefont
  {Paar}}, \bibinfo {author} {\bibfnamefont {Dario}\ \bibnamefont {Vretenar}},
  \bibinfo {author} {\bibfnamefont {Elias}\ \bibnamefont {Khan}}, \ and\
  \bibinfo {author} {\bibfnamefont {Gianluca}\ \bibnamefont {Colò}},\
  }\bibfield  {title} {\enquote {\bibinfo {title} {Exotic modes of excitation
  in atomic nuclei far from stability},}\ }\href {\doibase
  10.1088/0034-4885/70/5/R02} {\bibfield  {journal} {\bibinfo  {journal}
  {Reports on Progress in Physics}\ }\textbf {\bibinfo {volume} {70}},\
  \bibinfo {pages} {R02} (\bibinfo {year} {2007})}\BibitemShut {NoStop}%
\bibitem [{\citenamefont {Colò}\ and\ \citenamefont
  {Roca-Maza}(2021)}]{Col2021}%
  \BibitemOpen
  \bibfield  {author} {\bibinfo {author} {\bibfnamefont {Gianluca}\
  \bibnamefont {Colò}}\ and\ \bibinfo {author} {\bibfnamefont {Xavier}\
  \bibnamefont {Roca-Maza}},\ }\href {https://arxiv.org/abs/2102.06562}
  {\enquote {\bibinfo {title} {User guide for the hfbcs-qrpa(v1) code},}\ }
  (\bibinfo {year} {2021}),\ \Eprint {http://arxiv.org/abs/2102.06562}
  {arXiv:2102.06562 [nucl-th]} \BibitemShut {NoStop}%
\bibitem [{\citenamefont {Fracasso}\ and\ \citenamefont
  {Col\`o}(2005)}]{Fra2005}%
  \BibitemOpen
  \bibfield  {author} {\bibinfo {author} {\bibfnamefont {S.}~\bibnamefont
  {Fracasso}}\ and\ \bibinfo {author} {\bibfnamefont {G.}~\bibnamefont
  {Col\`o}},\ }\bibfield  {title} {\enquote {\bibinfo {title} {Fully
  self-consistent charge-exchange quasiparticle random-phase approximation and
  its application to isobaric analog resonances},}\ }\href {\doibase
  10.1103/PhysRevC.72.064310} {\bibfield  {journal} {\bibinfo  {journal} {Phys.
  Rev. C}\ }\textbf {\bibinfo {volume} {72}},\ \bibinfo {pages} {064310}
  (\bibinfo {year} {2005})}\BibitemShut {NoStop}%
\bibitem [{\citenamefont {Roca-Maza}\ \emph {et~al.}(2012)\citenamefont
  {Roca-Maza}, \citenamefont {Col\`o},\ and\ \citenamefont
  {Sagawa}}]{RocaMaza2012}%
  \BibitemOpen
  \bibfield  {author} {\bibinfo {author} {\bibfnamefont {X.}~\bibnamefont
  {Roca-Maza}}, \bibinfo {author} {\bibfnamefont {G.}~\bibnamefont {Col\`o}}, \
  and\ \bibinfo {author} {\bibfnamefont {H.}~\bibnamefont {Sagawa}},\
  }\bibfield  {title} {\enquote {\bibinfo {title} {New {Skyrme} interaction
  with improved spin-isospin properties},}\ }\href {\doibase
  10.1103/PhysRevC.86.031306} {\bibfield  {journal} {\bibinfo  {journal} {Phys.
  Rev. C}\ }\textbf {\bibinfo {volume} {86}},\ \bibinfo {pages} {031306}
  (\bibinfo {year} {2012})}\BibitemShut {NoStop}%
\bibitem [{\citenamefont {Auerbach}\ and\ \citenamefont
  {Klein}(1983)}]{Auerbach83}%
  \BibitemOpen
  \bibfield  {author} {\bibinfo {author} {\bibfnamefont {N.}~\bibnamefont
  {Auerbach}}\ and\ \bibinfo {author} {\bibfnamefont {A.}~\bibnamefont
  {Klein}},\ }\bibfield  {title} {\enquote {\bibinfo {title} {A microscopic
  theory of giant electric isovector resonances},}\ }\href {\doibase
  https://doi.org/10.1016/0375-9474(83)90090-8} {\bibfield  {journal} {\bibinfo
   {journal} {Nuclear Physics A}\ }\textbf {\bibinfo {volume} {395}},\ \bibinfo
  {pages} {77--118} (\bibinfo {year} {1983})}\BibitemShut {NoStop}%
\bibitem [{\citenamefont {Leonardi}(1976)}]{Leonardi1976}%
  \BibitemOpen
  \bibfield  {author} {\bibinfo {author} {\bibfnamefont {Renzo}\ \bibnamefont
  {Leonardi}},\ }\bibfield  {title} {\enquote {\bibinfo {title} {Effects of the
  isotensor potential in nuclear excitation carrying isospin},}\ }\href
  {\doibase 10.1103/PhysRevC.14.385} {\bibfield  {journal} {\bibinfo  {journal}
  {Phys. Rev. C}\ }\textbf {\bibinfo {volume} {14}},\ \bibinfo {pages}
  {385--401} (\bibinfo {year} {1976})}\BibitemShut {NoStop}%
\bibitem [{\citenamefont {Lipparini}\ and\ \citenamefont
  {Stringari}(1989)}]{lipparini1989}%
  \BibitemOpen
  \bibfield  {author} {\bibinfo {author} {\bibfnamefont {E.}~\bibnamefont
  {Lipparini}}\ and\ \bibinfo {author} {\bibfnamefont {S.}~\bibnamefont
  {Stringari}},\ }\bibfield  {title} {\enquote {\bibinfo {title} {Sum rules and
  giant resonances in nuclei},}\ }\href {\doibase
  https://doi.org/10.1016/0370-1573(89)90029-X} {\bibfield  {journal} {\bibinfo
   {journal} {Physics Reports}\ }\textbf {\bibinfo {volume} {175}},\ \bibinfo
  {pages} {103--261} (\bibinfo {year} {1989})}\BibitemShut {NoStop}%
\bibitem [{\citenamefont {Bohr}\ and\ \citenamefont {Mottelson}(1998)}]{bm}%
  \BibitemOpen
  \bibfield  {author} {\bibinfo {author} {\bibfnamefont {Aage}\ \bibnamefont
  {Bohr}}\ and\ \bibinfo {author} {\bibfnamefont {Ben~R.}\ \bibnamefont
  {Mottelson}},\ }\href@noop {} {\emph {\bibinfo {title} {Nuclear
  Structure}}},\ Vol.\ \bibinfo {volume} {I \& II}\ (\bibinfo  {publisher}
  {World Scientific Publishing Company},\ \bibinfo {address} {Singapore},\
  \bibinfo {year} {1998})\BibitemShut {NoStop}%
\bibitem [{\citenamefont {Centelles}\ \emph {et~al.}(2009)\citenamefont
  {Centelles}, \citenamefont {Roca-Maza}, \citenamefont {Vi\~nas},\ and\
  \citenamefont {Warda}}]{Centelles2009}%
  \BibitemOpen
  \bibfield  {author} {\bibinfo {author} {\bibfnamefont {M.}~\bibnamefont
  {Centelles}}, \bibinfo {author} {\bibfnamefont {X.}~\bibnamefont
  {Roca-Maza}}, \bibinfo {author} {\bibfnamefont {X.}~\bibnamefont {Vi\~nas}},
  \ and\ \bibinfo {author} {\bibfnamefont {M.}~\bibnamefont {Warda}},\
  }\bibfield  {title} {\enquote {\bibinfo {title} {Nuclear symmetry energy
  probed by neutron skin thickness of nuclei},}\ }\href {\doibase
  10.1103/PhysRevLett.102.122502} {\bibfield  {journal} {\bibinfo  {journal}
  {Phys. Rev. Lett.}\ }\textbf {\bibinfo {volume} {102}},\ \bibinfo {pages}
  {122502} (\bibinfo {year} {2009})}\BibitemShut {NoStop}%
\bibitem [{\citenamefont {Bohigas}\ \emph {et~al.}(1979)\citenamefont
  {Bohigas}, \citenamefont {Lane},\ and\ \citenamefont
  {Martorell}}]{bohigas1979}%
  \BibitemOpen
  \bibfield  {author} {\bibinfo {author} {\bibfnamefont {O.}~\bibnamefont
  {Bohigas}}, \bibinfo {author} {\bibfnamefont {A.M.}\ \bibnamefont {Lane}}, \
  and\ \bibinfo {author} {\bibfnamefont {J.}~\bibnamefont {Martorell}},\
  }\bibfield  {title} {\enquote {\bibinfo {title} {Sum rules for nuclear
  collective excitations},}\ }\href {\doibase
  https://doi.org/10.1016/0370-1573(79)90079-6} {\bibfield  {journal} {\bibinfo
   {journal} {Physics Reports}\ }\textbf {\bibinfo {volume} {51}},\ \bibinfo
  {pages} {267--316} (\bibinfo {year} {1979})}\BibitemShut {NoStop}%
\bibitem [{\citenamefont {Capelli}\ \emph {et~al.}(2009)\citenamefont
  {Capelli}, \citenamefont {Col{\`o}},\ and\ \citenamefont {Li}}]{capelli2009}%
  \BibitemOpen
  \bibfield  {author} {\bibinfo {author} {\bibfnamefont {Luigi}\ \bibnamefont
  {Capelli}}, \bibinfo {author} {\bibfnamefont {Gianluca}\ \bibnamefont
  {Col{\`o}}}, \ and\ \bibinfo {author} {\bibfnamefont {Jun}\ \bibnamefont
  {Li}},\ }\bibfield  {title} {\enquote {\bibinfo {title} {Dielectric theorem
  within the hartree-fock-bogoliubov framework},}\ }\href {\doibase
  10.1103/PhysRevC.79.054329} {\bibfield  {journal} {\bibinfo  {journal} {Phys.
  Rev. C}\ }\textbf {\bibinfo {volume} {79}},\ \bibinfo {pages} {054329}
  (\bibinfo {year} {2009})}\BibitemShut {NoStop}%
\bibitem [{\citenamefont {Hinohara}\ \emph {et~al.}(2015)\citenamefont
  {Hinohara}, \citenamefont {Kortelainen}, \citenamefont {Nazarewicz},\ and\
  \citenamefont {Olsen}}]{Hinohara2015}%
  \BibitemOpen
  \bibfield  {author} {\bibinfo {author} {\bibfnamefont {Nobuo}\ \bibnamefont
  {Hinohara}}, \bibinfo {author} {\bibfnamefont {Markus}\ \bibnamefont
  {Kortelainen}}, \bibinfo {author} {\bibfnamefont {Witold}\ \bibnamefont
  {Nazarewicz}}, \ and\ \bibinfo {author} {\bibfnamefont {Erik}\ \bibnamefont
  {Olsen}},\ }\bibfield  {title} {\enquote {\bibinfo {title} {Complex-energy
  approach to sum rules within nuclear density functional theory},}\ }\href
  {\doibase 10.1103/PhysRevC.91.044323} {\bibfield  {journal} {\bibinfo
  {journal} {Phys. Rev. C}\ }\textbf {\bibinfo {volume} {91}},\ \bibinfo
  {pages} {044323} (\bibinfo {year} {2015})}\BibitemShut {NoStop}%
\bibitem [{\citenamefont {Krivine}\ \emph {et~al.}(1980)\citenamefont
  {Krivine}, \citenamefont {Treiner},\ and\ \citenamefont
  {Bohigas}}]{kirvine1980}%
  \BibitemOpen
  \bibfield  {author} {\bibinfo {author} {\bibfnamefont {H.}~\bibnamefont
  {Krivine}}, \bibinfo {author} {\bibfnamefont {J.}~\bibnamefont {Treiner}}, \
  and\ \bibinfo {author} {\bibfnamefont {O.}~\bibnamefont {Bohigas}},\
  }\bibfield  {title} {\enquote {\bibinfo {title} {Derivation of a
  fluid-dynamical lagrangian and electric giant resonances},}\ }\href {\doibase
  https://doi.org/10.1016/0375-9474(80)90618-1} {\bibfield  {journal} {\bibinfo
   {journal} {Nuclear Physics A}\ }\textbf {\bibinfo {volume} {336}},\ \bibinfo
  {pages} {155--184} (\bibinfo {year} {1980})}\BibitemShut {NoStop}%
\bibitem [{\citenamefont {Casas}\ \emph {et~al.}(1985)\citenamefont {Casas},
  \citenamefont {Martorell},\ and\ \citenamefont {Gomez}}]{casas1985}%
  \BibitemOpen
  \bibfield  {author} {\bibinfo {author} {\bibfnamefont {M.}~\bibnamefont
  {Casas}}, \bibinfo {author} {\bibfnamefont {J.}~\bibnamefont {Martorell}}, \
  and\ \bibinfo {author} {\bibfnamefont {J.M.G.}\ \bibnamefont {Gomez}},\
  }\bibfield  {title} {\enquote {\bibinfo {title} {Giant isovector monopole
  resonances and the symmetry energy of nuclear matter},}\ }\href {\doibase
  https://doi.org/10.1016/0370-2693(85)91128-1} {\bibfield  {journal} {\bibinfo
   {journal} {Physics Letters B}\ }\textbf {\bibinfo {volume} {152}},\ \bibinfo
  {pages} {6--10} (\bibinfo {year} {1985})}\BibitemShut {NoStop}%
\bibitem [{\citenamefont {Colò}\ \emph {et~al.}(1995)\citenamefont {Colò},
  \citenamefont {{Nguyen Van Giai}},\ and\ \citenamefont {Sagawa}}]{Colo1995}%
  \BibitemOpen
  \bibfield  {author} {\bibinfo {author} {\bibfnamefont {G.}~\bibnamefont
  {Colò}}, \bibinfo {author} {\bibnamefont {{Nguyen Van Giai}}}, \ and\
  \bibinfo {author} {\bibfnamefont {H.}~\bibnamefont {Sagawa}},\ }\bibfield
  {title} {\enquote {\bibinfo {title} {Isovector properties of skyrme-type
  effective interactions},}\ }\href {\doibase
  https://doi.org/10.1016/0370-2693(95)01217-E} {\bibfield  {journal} {\bibinfo
   {journal} {Physics Letters B}\ }\textbf {\bibinfo {volume} {363}},\ \bibinfo
  {pages} {5--11} (\bibinfo {year} {1995})}\BibitemShut {NoStop}%
\bibitem [{\citenamefont {Myers}\ and\ \citenamefont
  {Swiatecki}(1974)}]{myers1974}%
  \BibitemOpen
  \bibfield  {author} {\bibinfo {author} {\bibfnamefont {W.D}\ \bibnamefont
  {Myers}}\ and\ \bibinfo {author} {\bibfnamefont {W.J}\ \bibnamefont
  {Swiatecki}},\ }\bibfield  {title} {\enquote {\bibinfo {title} {The nuclear
  droplet model for arbitrary shapes},}\ }\href {\doibase
  https://doi.org/10.1016/0003-4916(74)90299-1} {\bibfield  {journal} {\bibinfo
   {journal} {Annals of Physics}\ }\textbf {\bibinfo {volume} {84}},\ \bibinfo
  {pages} {186--210} (\bibinfo {year} {1974})}\BibitemShut {NoStop}%
\bibitem [{\citenamefont {Meyer}\ \emph {et~al.}(1982)\citenamefont {Meyer},
  \citenamefont {Quentin},\ and\ \citenamefont {Jennings}}]{meyer1982}%
  \BibitemOpen
  \bibfield  {author} {\bibinfo {author} {\bibfnamefont {J.}~\bibnamefont
  {Meyer}}, \bibinfo {author} {\bibfnamefont {P.}~\bibnamefont {Quentin}}, \
  and\ \bibinfo {author} {\bibfnamefont {B.K.}\ \bibnamefont {Jennings}},\
  }\bibfield  {title} {\enquote {\bibinfo {title} {The isovector dipole mode: A
  simple sum rule approach},}\ }\href {\doibase
  https://doi.org/10.1016/0375-9474(82)90172-5} {\bibfield  {journal} {\bibinfo
   {journal} {Nuclear Physics A}\ }\textbf {\bibinfo {volume} {385}},\ \bibinfo
  {pages} {269--284} (\bibinfo {year} {1982})}\BibitemShut {NoStop}%
\bibitem [{\citenamefont {Shen}\ \emph {et~al.}(2019)\citenamefont {Shen},
  \citenamefont {Col\`o},\ and\ \citenamefont {Roca-Maza}}]{Shen2019}%
  \BibitemOpen
  \bibfield  {author} {\bibinfo {author} {\bibfnamefont {Shihang}\ \bibnamefont
  {Shen}}, \bibinfo {author} {\bibfnamefont {Gianluca}\ \bibnamefont {Col\`o}},
  \ and\ \bibinfo {author} {\bibfnamefont {Xavier}\ \bibnamefont {Roca-Maza}},\
  }\bibfield  {title} {\enquote {\bibinfo {title} {Skyrme functional with
  tensor terms from ab initio calculations of neutron-proton drops},}\ }\href
  {\doibase 10.1103/PhysRevC.99.034322} {\bibfield  {journal} {\bibinfo
  {journal} {Phys. Rev. C}\ }\textbf {\bibinfo {volume} {99}},\ \bibinfo
  {pages} {034322} (\bibinfo {year} {2019})}\BibitemShut {NoStop}%
\bibitem [{\citenamefont {Roca-Maza}\ \emph
  {et~al.}(2013{\natexlab{a}})\citenamefont {Roca-Maza}, \citenamefont
  {Brenna}, \citenamefont {Agrawal}, \citenamefont {Bortignon}, \citenamefont
  {Col\`o}, \citenamefont {Cao}, \citenamefont {Paar},\ and\ \citenamefont
  {Vretenar}}]{Roca-Maza2013}%
  \BibitemOpen
  \bibfield  {author} {\bibinfo {author} {\bibfnamefont {X.}~\bibnamefont
  {Roca-Maza}}, \bibinfo {author} {\bibfnamefont {M.}~\bibnamefont {Brenna}},
  \bibinfo {author} {\bibfnamefont {B.~K.}\ \bibnamefont {Agrawal}}, \bibinfo
  {author} {\bibfnamefont {P.~F.}\ \bibnamefont {Bortignon}}, \bibinfo {author}
  {\bibfnamefont {G.}~\bibnamefont {Col\`o}}, \bibinfo {author} {\bibfnamefont
  {Li-Gang}\ \bibnamefont {Cao}}, \bibinfo {author} {\bibfnamefont
  {N.}~\bibnamefont {Paar}}, \ and\ \bibinfo {author} {\bibfnamefont
  {D.}~\bibnamefont {Vretenar}},\ }\bibfield  {title} {\enquote {\bibinfo
  {title} {Giant quadrupole resonances in ${}^{208}$pb, the nuclear symmetry
  energy, and the neutron skin thickness},}\ }\href {\doibase
  10.1103/PhysRevC.87.034301} {\bibfield  {journal} {\bibinfo  {journal} {Phys.
  Rev. C}\ }\textbf {\bibinfo {volume} {87}},\ \bibinfo {pages} {034301}
  (\bibinfo {year} {2013}{\natexlab{a}})}\BibitemShut {NoStop}%
\bibitem [{\citenamefont {Agrawal}\ \emph {et~al.}(2005)\citenamefont
  {Agrawal}, \citenamefont {Shlomo},\ and\ \citenamefont {Au}}]{Agrawal2005}%
  \BibitemOpen
  \bibfield  {author} {\bibinfo {author} {\bibfnamefont {B.~K.}\ \bibnamefont
  {Agrawal}}, \bibinfo {author} {\bibfnamefont {S.}~\bibnamefont {Shlomo}}, \
  and\ \bibinfo {author} {\bibfnamefont {V.~Kim}\ \bibnamefont {Au}},\
  }\bibfield  {title} {\enquote {\bibinfo {title} {Determination of the
  parameters of a skyrme type effective interaction using the simulated
  annealing approach},}\ }\href {\doibase 10.1103/PhysRevC.72.014310}
  {\bibfield  {journal} {\bibinfo  {journal} {Phys. Rev. C}\ }\textbf {\bibinfo
  {volume} {72}},\ \bibinfo {pages} {014310} (\bibinfo {year}
  {2005})}\BibitemShut {NoStop}%
\bibitem [{\citenamefont {{Van Giai}}\ and\ \citenamefont
  {Sagawa}(1981)}]{Giai1981}%
  \BibitemOpen
  \bibfield  {author} {\bibinfo {author} {\bibfnamefont {Nguyen}\ \bibnamefont
  {{Van Giai}}}\ and\ \bibinfo {author} {\bibfnamefont {H.}~\bibnamefont
  {Sagawa}},\ }\bibfield  {title} {\enquote {\bibinfo {title} {Spin-isospin and
  pairing properties of modified skyrme interactions},}\ }\href {\doibase
  https://doi.org/10.1016/0370-2693(81)90646-8} {\bibfield  {journal} {\bibinfo
   {journal} {Physics Letters B}\ }\textbf {\bibinfo {volume} {106}},\ \bibinfo
  {pages} {379--382} (\bibinfo {year} {1981})}\BibitemShut {NoStop}%
\bibitem [{\citenamefont {Beiner}\ \emph {et~al.}(1975)\citenamefont {Beiner},
  \citenamefont {Flocard}, \citenamefont {{Van Giai}},\ and\ \citenamefont
  {Quentin}}]{Beiner1975}%
  \BibitemOpen
  \bibfield  {author} {\bibinfo {author} {\bibfnamefont {M.}~\bibnamefont
  {Beiner}}, \bibinfo {author} {\bibfnamefont {H.}~\bibnamefont {Flocard}},
  \bibinfo {author} {\bibfnamefont {Nguyen}\ \bibnamefont {{Van Giai}}}, \ and\
  \bibinfo {author} {\bibfnamefont {P.}~\bibnamefont {Quentin}},\ }\bibfield
  {title} {\enquote {\bibinfo {title} {Nuclear ground-state properties and
  self-consistent calculations with the skyrme interaction: (i). spherical
  description},}\ }\href {\doibase
  https://doi.org/10.1016/0375-9474(75)90338-3} {\bibfield  {journal} {\bibinfo
   {journal} {Nuclear Physics A}\ }\textbf {\bibinfo {volume} {238}},\ \bibinfo
  {pages} {29--69} (\bibinfo {year} {1975})}\BibitemShut {NoStop}%
\bibitem [{\citenamefont {Reinhard}\ and\ \citenamefont
  {Flocard}(1995)}]{Reinhard1995}%
  \BibitemOpen
  \bibfield  {author} {\bibinfo {author} {\bibfnamefont {P.-G.}\ \bibnamefont
  {Reinhard}}\ and\ \bibinfo {author} {\bibfnamefont {H.}~\bibnamefont
  {Flocard}},\ }\bibfield  {title} {\enquote {\bibinfo {title} {Nuclear
  effective forces and isotope shifts},}\ }\href {\doibase
  https://doi.org/10.1016/0375-9474(94)00770-N} {\bibfield  {journal} {\bibinfo
   {journal} {Nuclear Physics A}\ }\textbf {\bibinfo {volume} {584}},\ \bibinfo
  {pages} {467--488} (\bibinfo {year} {1995})}\BibitemShut {NoStop}%
\bibitem [{\citenamefont {Bartel}\ \emph {et~al.}(1982)\citenamefont {Bartel},
  \citenamefont {Quentin}, \citenamefont {Brack}, \citenamefont {Guet},\ and\
  \citenamefont {Håkansson}}]{Bartel1982}%
  \BibitemOpen
  \bibfield  {author} {\bibinfo {author} {\bibfnamefont {J.}~\bibnamefont
  {Bartel}}, \bibinfo {author} {\bibfnamefont {P.}~\bibnamefont {Quentin}},
  \bibinfo {author} {\bibfnamefont {M.}~\bibnamefont {Brack}}, \bibinfo
  {author} {\bibfnamefont {C.}~\bibnamefont {Guet}}, \ and\ \bibinfo {author}
  {\bibfnamefont {H.-B.}\ \bibnamefont {Håkansson}},\ }\bibfield  {title}
  {\enquote {\bibinfo {title} {Towards a better parametrisation of skyrme-like
  effective forces: A critical study of the skm force},}\ }\href {\doibase
  https://doi.org/10.1016/0375-9474(82)90403-1} {\bibfield  {journal} {\bibinfo
   {journal} {Nuclear Physics A}\ }\textbf {\bibinfo {volume} {386}},\ \bibinfo
  {pages} {79--100} (\bibinfo {year} {1982})}\BibitemShut {NoStop}%
\bibitem [{\citenamefont {Reinhard}\ \emph {et~al.}(1999)\citenamefont
  {Reinhard}, \citenamefont {Dean}, \citenamefont {Nazarewicz}, \citenamefont
  {Dobaczewski}, \citenamefont {Maruhn},\ and\ \citenamefont
  {Strayer}}]{Reinhard1999}%
  \BibitemOpen
  \bibfield  {author} {\bibinfo {author} {\bibfnamefont {P.-G.}\ \bibnamefont
  {Reinhard}}, \bibinfo {author} {\bibfnamefont {D.~J.}\ \bibnamefont {Dean}},
  \bibinfo {author} {\bibfnamefont {W.}~\bibnamefont {Nazarewicz}}, \bibinfo
  {author} {\bibfnamefont {J.}~\bibnamefont {Dobaczewski}}, \bibinfo {author}
  {\bibfnamefont {J.~A.}\ \bibnamefont {Maruhn}}, \ and\ \bibinfo {author}
  {\bibfnamefont {M.~R.}\ \bibnamefont {Strayer}},\ }\bibfield  {title}
  {\enquote {\bibinfo {title} {Shape coexistence and the effective
  nucleon-nucleon interaction},}\ }\href {\doibase 10.1103/PhysRevC.60.014316}
  {\bibfield  {journal} {\bibinfo  {journal} {Phys. Rev. C}\ }\textbf {\bibinfo
  {volume} {60}},\ \bibinfo {pages} {014316} (\bibinfo {year}
  {1999})}\BibitemShut {NoStop}%
\bibitem [{\citenamefont {Colò}\ \emph {et~al.}(2007)\citenamefont {Colò},
  \citenamefont {Sagawa}, \citenamefont {Fracasso},\ and\ \citenamefont
  {Bortignon}}]{Colo2007}%
  \BibitemOpen
  \bibfield  {author} {\bibinfo {author} {\bibfnamefont {G.}~\bibnamefont
  {Colò}}, \bibinfo {author} {\bibfnamefont {H.}~\bibnamefont {Sagawa}},
  \bibinfo {author} {\bibfnamefont {S.}~\bibnamefont {Fracasso}}, \ and\
  \bibinfo {author} {\bibfnamefont {P.F.}\ \bibnamefont {Bortignon}},\
  }\bibfield  {title} {\enquote {\bibinfo {title} {Spin–orbit splitting and
  the tensor component of the skyrme interaction},}\ }\href {\doibase
  https://doi.org/10.1016/j.physletb.2007.01.033} {\bibfield  {journal}
  {\bibinfo  {journal} {Physics Letters B}\ }\textbf {\bibinfo {volume}
  {646}},\ \bibinfo {pages} {227--231} (\bibinfo {year} {2007})}\BibitemShut
  {NoStop}%
\bibitem [{\citenamefont {Chabanat}\ \emph {et~al.}(1998)\citenamefont
  {Chabanat}, \citenamefont {Bonche}, \citenamefont {Haensel}, \citenamefont
  {Meyer},\ and\ \citenamefont {Schaeffer}}]{Chabanat1998}%
  \BibitemOpen
  \bibfield  {author} {\bibinfo {author} {\bibfnamefont {E.}~\bibnamefont
  {Chabanat}}, \bibinfo {author} {\bibfnamefont {P.}~\bibnamefont {Bonche}},
  \bibinfo {author} {\bibfnamefont {P.}~\bibnamefont {Haensel}}, \bibinfo
  {author} {\bibfnamefont {J.}~\bibnamefont {Meyer}}, \ and\ \bibinfo {author}
  {\bibfnamefont {R.}~\bibnamefont {Schaeffer}},\ }\bibfield  {title} {\enquote
  {\bibinfo {title} {A skyrme parametrization from subnuclear to neutron star
  densities part ii. nuclei far from stabilities},}\ }\href {\doibase
  https://doi.org/10.1016/S0375-9474(98)00180-8} {\bibfield  {journal}
  {\bibinfo  {journal} {Nuclear Physics A}\ }\textbf {\bibinfo {volume}
  {635}},\ \bibinfo {pages} {231--256} (\bibinfo {year} {1998})}\BibitemShut
  {NoStop}%
\bibitem [{\citenamefont {Agrawal}\ \emph {et~al.}(2003)\citenamefont
  {Agrawal}, \citenamefont {Shlomo},\ and\ \citenamefont
  {Kim~Au}}]{Agrawal2003}%
  \BibitemOpen
  \bibfield  {author} {\bibinfo {author} {\bibfnamefont {B.~K.}\ \bibnamefont
  {Agrawal}}, \bibinfo {author} {\bibfnamefont {S.}~\bibnamefont {Shlomo}}, \
  and\ \bibinfo {author} {\bibfnamefont {V.}~\bibnamefont {Kim~Au}},\
  }\bibfield  {title} {\enquote {\bibinfo {title} {Nuclear matter
  incompressibility coefficient in relativistic and nonrelativistic microscopic
  models},}\ }\href {\doibase 10.1103/PhysRevC.68.031304} {\bibfield  {journal}
  {\bibinfo  {journal} {Phys. Rev. C}\ }\textbf {\bibinfo {volume} {68}},\
  \bibinfo {pages} {031304} (\bibinfo {year} {2003})}\BibitemShut {NoStop}%
\bibitem [{\citenamefont {Kl\"upfel}\ \emph {et~al.}(2009)\citenamefont
  {Kl\"upfel}, \citenamefont {Reinhard}, \citenamefont {B\"urvenich},\ and\
  \citenamefont {Maruhn}}]{kupfel2009}%
  \BibitemOpen
  \bibfield  {author} {\bibinfo {author} {\bibfnamefont {P.}~\bibnamefont
  {Kl\"upfel}}, \bibinfo {author} {\bibfnamefont {P.-G.}\ \bibnamefont
  {Reinhard}}, \bibinfo {author} {\bibfnamefont {T.~J.}\ \bibnamefont
  {B\"urvenich}}, \ and\ \bibinfo {author} {\bibfnamefont {J.~A.}\ \bibnamefont
  {Maruhn}},\ }\bibfield  {title} {\enquote {\bibinfo {title} {Variations on a
  theme by skyrme: A systematic study of adjustments of model parameters},}\
  }\href {\doibase 10.1103/PhysRevC.79.034310} {\bibfield  {journal} {\bibinfo
  {journal} {Phys. Rev. C}\ }\textbf {\bibinfo {volume} {79}},\ \bibinfo
  {pages} {034310} (\bibinfo {year} {2009})}\BibitemShut {NoStop}%
\bibitem [{\citenamefont {Dobaczewski}\ \emph {et~al.}(1984)\citenamefont
  {Dobaczewski}, \citenamefont {Flocard},\ and\ \citenamefont
  {Treiner}}]{Dobaczewski1984}%
  \BibitemOpen
  \bibfield  {author} {\bibinfo {author} {\bibfnamefont {J.}~\bibnamefont
  {Dobaczewski}}, \bibinfo {author} {\bibfnamefont {H.}~\bibnamefont
  {Flocard}}, \ and\ \bibinfo {author} {\bibfnamefont {J.}~\bibnamefont
  {Treiner}},\ }\bibfield  {title} {\enquote {\bibinfo {title}
  {Hartree-fock-bogolyubov description of nuclei near the neutron-drip line},}\
  }\href {\doibase https://doi.org/10.1016/0375-9474(84)90433-0} {\bibfield
  {journal} {\bibinfo  {journal} {Nuclear Physics A}\ }\textbf {\bibinfo
  {volume} {422}},\ \bibinfo {pages} {103--139} (\bibinfo {year}
  {1984})}\BibitemShut {NoStop}%
\bibitem [{\citenamefont {Alex~Brown}(1998)}]{Brown1998}%
  \BibitemOpen
  \bibfield  {author} {\bibinfo {author} {\bibfnamefont {B.}~\bibnamefont
  {Alex~Brown}},\ }\bibfield  {title} {\enquote {\bibinfo {title} {New skyrme
  interaction for normal and exotic nuclei},}\ }\href {\doibase
  10.1103/PhysRevC.58.220} {\bibfield  {journal} {\bibinfo  {journal} {Phys.
  Rev. C}\ }\textbf {\bibinfo {volume} {58}},\ \bibinfo {pages} {220--231}
  (\bibinfo {year} {1998})}\BibitemShut {NoStop}%
\bibitem [{\citenamefont {Skovgaard}()}]{Skovgaard1998}%
  \BibitemOpen
  \bibfield  {author} {\bibinfo {author} {\bibfnamefont {L.T.}\ \bibnamefont
  {Skovgaard}},\ }\bibfield  {title} {\enquote {\bibinfo {title} {Applied
  regression analysis. 3rd edn. n. r. draper and h. smith, wiley, new york,
  1998. no. of pages: xvii+706. price: £45. isbn 0-471-17082-8},}\ }\href
  {\doibase
  https://doi.org/10.1002/1097-0258(20001130)19:22<3136::AID-SIM607>3.0.CO;2-Q}
  {\bibfield  {journal} {\bibinfo  {journal} {Statistics in Medicine}\ }\textbf
  {\bibinfo {volume} {19}},\ \bibinfo {pages} {3136--3139}}\BibitemShut
  {NoStop}%
\bibitem [{\citenamefont {Roca-Maza}\ \emph
  {et~al.}(2013{\natexlab{b}})\citenamefont {Roca-Maza}, \citenamefont
  {Brenna}, \citenamefont {Col{\`o}}, \citenamefont {Centelles}, \citenamefont
  {Vi{\~n}as}, \citenamefont {Agrawal}, \citenamefont {Paar}, \citenamefont
  {Vretenar},\ and\ \citenamefont {Piekarewicz}}]{Roca-Maza2013b}%
  \BibitemOpen
  \bibfield  {author} {\bibinfo {author} {\bibfnamefont {X.}~\bibnamefont
  {Roca-Maza}}, \bibinfo {author} {\bibfnamefont {M.}~\bibnamefont {Brenna}},
  \bibinfo {author} {\bibfnamefont {G.}~\bibnamefont {Col{\`o}}}, \bibinfo
  {author} {\bibfnamefont {M.}~\bibnamefont {Centelles}}, \bibinfo {author}
  {\bibfnamefont {X.}~\bibnamefont {Vi{\~n}as}}, \bibinfo {author}
  {\bibfnamefont {B.~K.}\ \bibnamefont {Agrawal}}, \bibinfo {author}
  {\bibfnamefont {N.}~\bibnamefont {Paar}}, \bibinfo {author} {\bibfnamefont
  {D.}~\bibnamefont {Vretenar}}, \ and\ \bibinfo {author} {\bibfnamefont
  {J.}~\bibnamefont {Piekarewicz}},\ }\bibfield  {title} {\enquote {\bibinfo
  {title} {Electric dipole polarizability in ${}^{208}$pb: Insights from the
  droplet model},}\ }\href {\doibase 10.1103/PhysRevC.88.024316} {\bibfield
  {journal} {\bibinfo  {journal} {Phys. Rev. C}\ }\textbf {\bibinfo {volume}
  {88}},\ \bibinfo {pages} {024316} (\bibinfo {year}
  {2013}{\natexlab{b}})}\BibitemShut {NoStop}%
\bibitem [{\citenamefont {Roca-Maza}\ \emph {et~al.}(2015)\citenamefont
  {Roca-Maza}, \citenamefont {Vi{\~n}as}, \citenamefont {Centelles},
  \citenamefont {Agrawal}, \citenamefont {Col{\`o}}, \citenamefont {Paar},
  \citenamefont {Piekarewicz},\ and\ \citenamefont {Vretenar}}]{Roca-Maza2015}%
  \BibitemOpen
  \bibfield  {author} {\bibinfo {author} {\bibfnamefont {X.}~\bibnamefont
  {Roca-Maza}}, \bibinfo {author} {\bibfnamefont {X.}~\bibnamefont
  {Vi{\~n}as}}, \bibinfo {author} {\bibfnamefont {M.}~\bibnamefont
  {Centelles}}, \bibinfo {author} {\bibfnamefont {B.~K.}\ \bibnamefont
  {Agrawal}}, \bibinfo {author} {\bibfnamefont {G.}~\bibnamefont {Col{\`o}}},
  \bibinfo {author} {\bibfnamefont {N.}~\bibnamefont {Paar}}, \bibinfo {author}
  {\bibfnamefont {J.}~\bibnamefont {Piekarewicz}}, \ and\ \bibinfo {author}
  {\bibfnamefont {D.}~\bibnamefont {Vretenar}},\ }\bibfield  {title} {\enquote
  {\bibinfo {title} {Neutron skin thickness from the measured electric dipole
  polarizability in $^{68}\text{Ni}$, $^{120}\text{Sn}$, and
  $^{208}\text{Pb}$},}\ }\href {\doibase 10.1103/PhysRevC.92.064304} {\bibfield
   {journal} {\bibinfo  {journal} {Phys. Rev. C}\ }\textbf {\bibinfo {volume}
  {92}},\ \bibinfo {pages} {064304} (\bibinfo {year} {2015})}\BibitemShut
  {NoStop}%
\end{thebibliography}%

\end{document}